\documentstyle[12pt,epsfig,amsfonts]{article}
\renewcommand{\theequation}{\arabic{section}.\arabic{equation}}
\title{Interference in the radiation\\ of two point-like charges}
\author{\bf Yurij Yaremko\footnote{Electronic mail:
yar@ph.icmp.lviv.ua}}
\date{\it Institute for Condensed Matter Physics, \\
1 Svientsitskii St., 79011 Lviv, Ukraine}
\begin{document}

\maketitle
\begin{abstract}
Energy-momentum and angular momentum carried by electromagnetic
field of two point-like charged particles arbitrarily moving in flat
spacetime are presented. Apart from usual contributions to the
Noether quantities produced separately by particles 1 and 2, the
conservation laws contain also joint contribution due to the fields
of both particles. The mixed part of Maxwell energy-momentum density
is decomposed into bound and radiative components which are
separately conserved off the world lines of particles. The former
describes the deformation of electromagnetic clouds of ``bare''
charges due to mutual interaction while the latter defines the
radiation which escapes to infinity. The bound terms contribute to
particles' individual 4-momenta while the radiative ones exert the
radiation reaction. Analysis of energy-momentum and angular momentum
balance equations results the Lorentz-Dirac equation as an equation
of motion for a pointed charge under the influence of its own
electromagnetic field as well as field produced by another charge.
\end{abstract}


\section{Introduction}\label{intro}
\setcounter{equation}{0}
The most natural and widely accepted equation of motion for a charge when
radiation reaction is taken into account is the Lorentz-Dirac equation
\cite{Dir}. This equation has been discussed mainly for the case of one
charge in an external electromagnetic field. In the present paper we
consider an isolated system of two point electric charges and their
electromagnetic field. We study the electromagnetic energy-momentum
and angular momentum radiated by charges; the study of energy-momentum and
angular momentum balance equations implies the Lorentz-Dirac equation for
more than one charge.

The dynamics of the entire system is governed by the action
\begin{equation}\label{S}
I=\sum\limits_{a=1}^2\left(-m_a\int {\rm d}s_a\sqrt{-{\dot
z}_a^2}+e_a\int {\rm d}s_a A_{a,\mu}{\dot z}_a^\mu\right)
-\frac{1}{16\pi}\int {\rm d}^4 y f_{\mu\nu}f^{\mu\nu},\nonumber
\end{equation}
where $f_{\mu\nu}=\sum_a(\partial_\mu A_{a,\nu}-\partial_\nu A_{a,\mu})$
is the total field generated by two charges. Charge $e_a$ moves on a
world line $\zeta_a\in{\mathbb M}_4$ described by functions
$z_a^\mu(s_a)$ which give the particle's coordinates as
functions of proper time $s_a$; ${\dot z}_a^\mu:=dz_a^\mu/ds_a$ is
the $a$-th four-velocity.

The action is invariant under the space-time translations and
rotations which constitute the Poincar\'e group. This immediately
implies conserved quantities which place stringent requirements on
the dynamics of the system. They demand that the change in
electromagnetic field momentum \cite{Rohr},
\begin{equation}\label{pem}
p^\nu_{\rm em}=\int_{\Sigma} {\rm d}\sigma_\alpha T^{\alpha\nu},
\end{equation}
and angular momentum \cite{Rohr},
\begin{equation}\label{Mem}
M^{\mu\nu}_{\rm em}=\int_{\Sigma} {\rm d}\sigma_\alpha\left(y^\mu
T^{\alpha\nu} - y^\nu T^{\alpha\mu}\right),
\end{equation}
should be balanced by a corresponding change in the total momentum and total
angular momentum of the particles. (By ${\rm d}\sigma_\alpha$ we denote the
vectorial surface element on a space-like hypersurface $\Sigma$.)

Since the Maxwell energy-momentum tensor density,
\begin{equation}\label{T}
4\pi T^{\mu\nu} = f^{\mu\lambda}f^\nu{}_\lambda - 1/4\eta^{\mu\nu}
f^{\kappa\lambda}f_{\kappa\lambda},
\end{equation}
is quadratic in the field and this field satisfies the superposition
principle, the total electromagnetic field stress-energy tensor is
\begin{equation}\label{TT}
T^{\mu\nu} = T_{(1)}^{\mu\nu}+T_{(2)}^{\mu\nu}+T_{\rm int}^{\mu\nu},
\end{equation}
where $a$-th particle density $T_{(a)}^{\mu\nu}$ is given by the expression
(\ref{T}) where ``total'' field strengths $f^{\mu\nu}$ are substituted by
``individual'' ones $f^{\mu\nu}_{(a)}$. The mixed term
\begin{equation}\label{T12}
4\pi T_{\rm int}^{\mu\nu} =
f_{(1)}^{\mu\lambda}f_{(2)}^\nu{}_\lambda +
f_{(2)}^{\mu\lambda}f_{(1)}^\nu{}_\lambda - 1/4\eta^{\mu\nu} \left(
f_{(1)}^{\kappa\lambda}f_{\kappa\lambda}^{(2)} +
f_{(2)}^{\kappa\lambda}f_{\kappa\lambda}^{(1)} \right)
\end{equation}
describes the joint contribution due to both fields.

In this paper we study radiation produced by an isolated system of
two point electric charges and their electromagnetic field. Outgoing
electromagnetic waves remove energy, momentum, and angular momentum
from the sources which then undergo radiation reaction. The
verification of conservation laws is not a trivial matter, since the
interference contribution (\ref{pint}) involves divergent terms.

In the derivation of particle's equation of motion, Dirac \cite{Dir}
evaluated the flux of electromagnetic energy-momentum over a narrow
world tube surrounding the particle's world line. The author
substituted the components of the retarded Li\'enard-Wiechert field
in the stress-energy tensor (\ref{T}) inside the narrow tube. In
1970 Teitelboim \cite{Teit} splits the ``retarded'' stress-energy
tensor into the ``bound'' and ``emitted'' parts which are separately
conserved off the world line of the particle. The author calculates
the flow of energy-momentum out of the portion of Bhabha world tube
\cite{Bha} bounded by tilted spacelike hypersurfaces which are
orthogonal to particle's four-velocity at instants $\tau$ and
$\tau+{\rm d}\tau$, respectively. Bound part, ${\hat T}_{\rm bnd}$,
describes a rigid electromagnetic ``cloud'' which are permanently
attached to the source and carried along with it. ``Bare'' charge
and ``cloud'' constitute new entity: dressed charged particle.
${\hat T}_{\rm bnd}$ contributes into particle's inertia: 4-momentum
of dressed charge contains, apart from usual velocity term, also a
term which is proportional to the square of charge $e_a$,
\begin{equation}\label{p_a}
p_a^\mu = m_au_a^\mu -\frac{2e_a^2}{3}a_a^\mu.
\end{equation}
(Coulomb-like infinity stemming from the pointness of ``bare''
source is absorbed by the rest mass $m_a$ within the renormalization
procedure.) Time derivative of the second term in eq.(\ref{p_a}) is
the well-known Schott term which describes a reversible form of
emission and absorption of field energy, which never gets far from
the point-like source. The radiative part, ${\hat T}_{\rm rad}$,
yields the Larmor relativistic rate of radiated energy-momentum. It
detaches itself from the charge and leads an independent existence.
This rate together with the Schott term constitutes the Abraham
radiation reaction vector. L\'opez and Villarroel \cite{LV} split
the torque of the stress-energy tensor into bound and emitted
components which possess analogous properties.

The results can be applied to the first and the second terms in the
total electromagnetic field stress-energy tensor (\ref{TT}) which
describe ``individual'' radiation contributions due to particles 1
and 2, respectively. The question is what part of the mixed density
should be taken instead of (\ref{T12}) to describe the radiation
which reaches to a very distant sphere?

Aguirregabiria and Bel \cite{AB} studied the interference part of
energy-momentum,
\begin{equation} \label{pint}
p_{\rm int}^\nu = \int_{\Sigma}d\sigma_\mu T_{\rm int}^{\mu\nu}\,,
\end{equation}
carried by electromagnetic field of two point charges. The authors
prove the fundamental theorem that the mixed radiation rate does not
depend on the shape of spacelike surface $\Sigma$ which is used to
integrate the mixed part (\ref{T12}) of the Maxwell energy-momentum
tensor density. For a prescribed plane motion of the charges the
perturbation scheme is elaborated within the framework of predictive
relativistic mechanics \cite{LM,LMM}. The lowest approximation gives
the well-known expression \cite[p.214]{LL} for the dipole radiation
of two point charges moving according to Coulomb's law. In Ref.
\cite{AE} the scheme is applied to the angular momentum carried by
electromagnetic field of two point charges.

In the case of $N$ particles we would merely obtain as an obvious
generalization of eq. (\ref{TT}) the sum of $N$ one-particle terms
and the mixed contributions corresponding to $N(N-1)/2$ pairs of
charges. Hence, the interference component dominates in the
radiation from a bunch of identical charged particles (e.g., in free
electron lasers). To evaluate the radiation of a relativistic
$N$-body system, Klepikov \cite{Kl} defines the center of a system
of radiation events which allows to synchronize the instants at
which electromagnetic waves emitted by different charges combine on
a very distant sphere. Fourier analysis is applied to calculate the
time and angular distributions of energy-momentum flux. The
radiation of a bunch of charged particles moving in a uniform
magnetic field is considered in detail.

The only exact solution is obtained by Rivera and Villarroel Ref.
\cite[eq.(3.27)]{VR}. The authors calculate the rate of radiation
(including interference part) generated by two identical point
charges rotating uniformly at opposite ends of a diameter, in a
fixed circle. External fields which govern the strictly prescribed
motions are constructed. The rate of radiation is evaluated via the
retarded Li\'enard-Wiechert fields produced by the charges. Further
\cite{RV} more general case of circular motion of two unlike charges
in two coplanar and concentric circumferences is considered.

Note that nine years before the radiation by a system of two uniformly
circling charges has been evaluated by Hnizdo \cite{Hn} (see also
discussion \cite{Hncom,RVrsp}). The author concludes that ``the power
radiated by such a system equals exactly the rate at which work is done on
the system by external force''.

In this paper we study the interference part of energy-momentum and
angular momentum of electromagnetic field generated by two
arbitrarily moving charges\footnote{The generalization of this work
to $N$ charges is an obvious one.}. We restrict ourselves to the
retarded Li\'enard-Wiechert solutions; the advanced ones are
rejected on the grounds of causality. In Section 2 we introduce
coordinate system which makes relatively easy the calculation of the
covariant 4-momentum radiated by interacting charges. The
calculation is performed in Sections 3 and 4. We reveal
divergence-free radiative part of the mixed density (\ref{T12}).  It
determines the radiation that escapes to infinity while the
(short-range) bound part modifies individual 4-momenta (\ref{p_a})
of dressed particles. The mixed part of radiated energy-momentum
depends only on velocities, accelerations and the relative
4-position of the charges. In Section 5 we derive equations of
motion of interacting charged particles. Analysis of energy-momentum
and angular momentum balance equations results the well-known
Lorentz-Dirac equation. In Section 6 we study symmetry properties of
radiative energy-momentum and angular momentum which rely on
invariance of action (\ref{S}) under inversions of space and time
axes. In Section 7 we discuss the results and implications.

\section{``Interference'' coordinate system}\label{coord}
\setcounter{equation}{0} To perform the surface integration
(\ref{pint}) of interference stress-energy tensor, an appropriate
coordinate system is necessary. Such a coordinate system is
introduced in Ref. \cite{AB}. It involves the evolution parameter
$\lambda$ associated with an inertial observer; the surface of
integration is a surface of constant $\lambda$. In Refs. \cite{Y1}
and \cite{Y2} this coordinate system is adapted to the simplest
hyperplane $\Sigma_t=\{y\in {\mathbb M}_{\,4}: y^0=t\}$ associated
with an unmoving inertial observer. The ``laboratory'' time $t$ is a
single common parameter defined along all the world lines of the
system.

\begin{figure}
\begin{center}
\epsfclipon
\epsfig{file=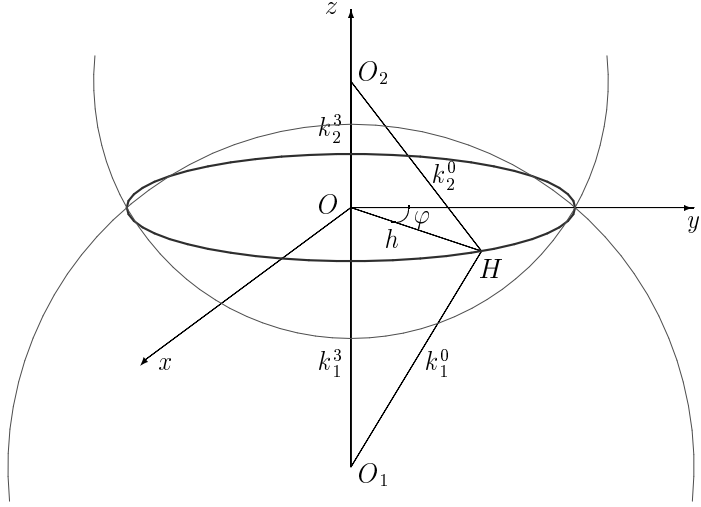,width=0.6\textwidth}
\end{center}
\caption{The interference picture in a plane $\Sigma_t$. Sphere $S_a$ with
radius $k_a^0=t-t_a$ is centered at point $O_a$ with coordinates
$z_a^{i'}(t_a), i'=1,2,3$. Angle $\varphi$ distinguishes the points of
intersection $S_1\cap S_2=C(O,h)$ which constitute support of integrals
(\ref{p_int}) and (\ref{M_int}). $k_a^0$, $k_a^3$, and $h$ are the
components of the future oriented null 4-vector
$k_a^{\alpha'}=\Omega^{\alpha'}{}_\alpha (y^\alpha-z_a^\alpha(t_a))$. Matrix
$\hat\Omega$ determines transition to ``momentarily rotating'' Lorentz frame.
}\label{kk}
\end{figure}

The mixed contributions to energy-momentum,
\begin{equation}\label{p_int}
p^\nu_{\rm int}(t)=\int_{\Sigma_t} {\rm d}\sigma_0 T_{\rm
int}^{0\nu},
\end{equation}
and angular momentum,
\begin{equation}\label{M_int}
M^{\mu\nu}_{\rm int}(t)=\int_{\Sigma_t} {\rm d}\sigma_0\left(y^\mu
T_{\rm int}^{0\nu} - y^\nu T_{\rm int}^{0\mu}\right),
\end{equation}
are due to interference of spherical wave fronts $S_1$ and $S_2$ in
$\Sigma_t$ (see Fig.~\ref{kk}). Sphere,
\begin{equation} \label{S_a}
S_a({\mathbf z}_a(t_a),t-t_a)=\{y\in
{\mathbb M}_{\,4}: (y^0-t_a)^2=\sum_i(y^i-z_a^i(t_a))^2,y^0=t,t-t_a>0\},
\end{equation}
is the intersection of the future light cone with vertex at point
$z_a(t_a)\in\zeta_a$ and $\Sigma_t$. This contribution is zero if
the relative position 4-vector $q=z_1-z_2$ is timelike. If $q$ is
spacelike, the intersection $S_1\cap S_2$ is the circle $C(O,h)$
with radius $h$; in ``momentarily rotating'' Lorentz frame the
circle $C(O,h)$ lies in $Oxy$ plane and centered at the coordinate
origin (see Fig.~\ref{kk}). If points $z_1$ and $z_2$ are related by
a null ray, the intersection $S_1\cap S_2$ contains the only point.

\subsection{Local map}
To find the local expressions for coordinate transformation
$(y^\alpha)\mapsto (t,t_1,t_2,\varphi)$, we translate the origin of
the laboratory Lorentz frame at the center $O$ of the circle
$C(O,h)=S_1\cap S_2$ and then rotate space axes till a new $z$-axis
be directed along 3-vector ${\mathbf q}:={\mathbf z}_1-{\mathbf
z}_2$,
\begin{equation} \label{Zha}
y^\alpha=z_a^\alpha(t_a)+\Omega^\alpha{}_{\alpha'}(t_1,t_2)k_a^{\alpha'}.
\end{equation}
Here $k_a, a=1,2$ is the future oriented null-vector with components
\begin{equation} \label{k_k}
k_a^0=t-t_a,\quad k_a^1=h\sin\varphi, \quad k_a^2=h\cos\varphi,
\quad k_a^3=(-1)^a\frac{\rm q}{2}+\frac{(k_2^0)^2-(k_1^0)^2}{2{\rm q}},
\end{equation}
which arise from analysis of triangle $O_1O_2H$ pictured in
Fig.~\ref{kk} (we denote ${\rm q}=|{\mathbf q}|$). Matrix space-time
components are $\Omega_{0\mu}=\Omega_{\mu 0}=\delta_{\mu 0}$. Its
space components $\Omega_{ij}$ constitute an orthogonal $3\times 3$
matrix (\ref{om}) which determines the rotation described above (see
\ref{varphi}). It defines new orthonormal basis,
\begin{eqnarray}\label{basis}
{\mathbf n}_\vartheta&=&\cos\varphi_q\cos\vartheta_q{\mathbf e}_1 +
\sin\varphi_q\cos\vartheta_q{\mathbf e}_2 -\sin\vartheta_q{\mathbf
e}_3\,,
\nonumber\\
{\mathbf n}_\varphi&=&-\sin\varphi_q{\mathbf e}_1+\cos\varphi_q{\mathbf
e}_2\,,\\
{\mathbf n}_q&=&\cos\varphi_q\sin\vartheta_q{\mathbf e}_1 +
\sin\varphi_q\sin\vartheta_q{\mathbf e}_2 +\cos\vartheta_q{\mathbf
e}_3\,, \nonumber
\end{eqnarray}
which is constructed from components of the relative position 3-vector
${\mathbf q}$, e.g. $\cos\varphi_q=q^1/\sqrt{(q^1)^2+(q^2)^2}$,
$\cos\vartheta_q=q^3/{\rm q}$.

\begin{figure}[t]
\begin{center}
\epsfclipon
\epsfig{file=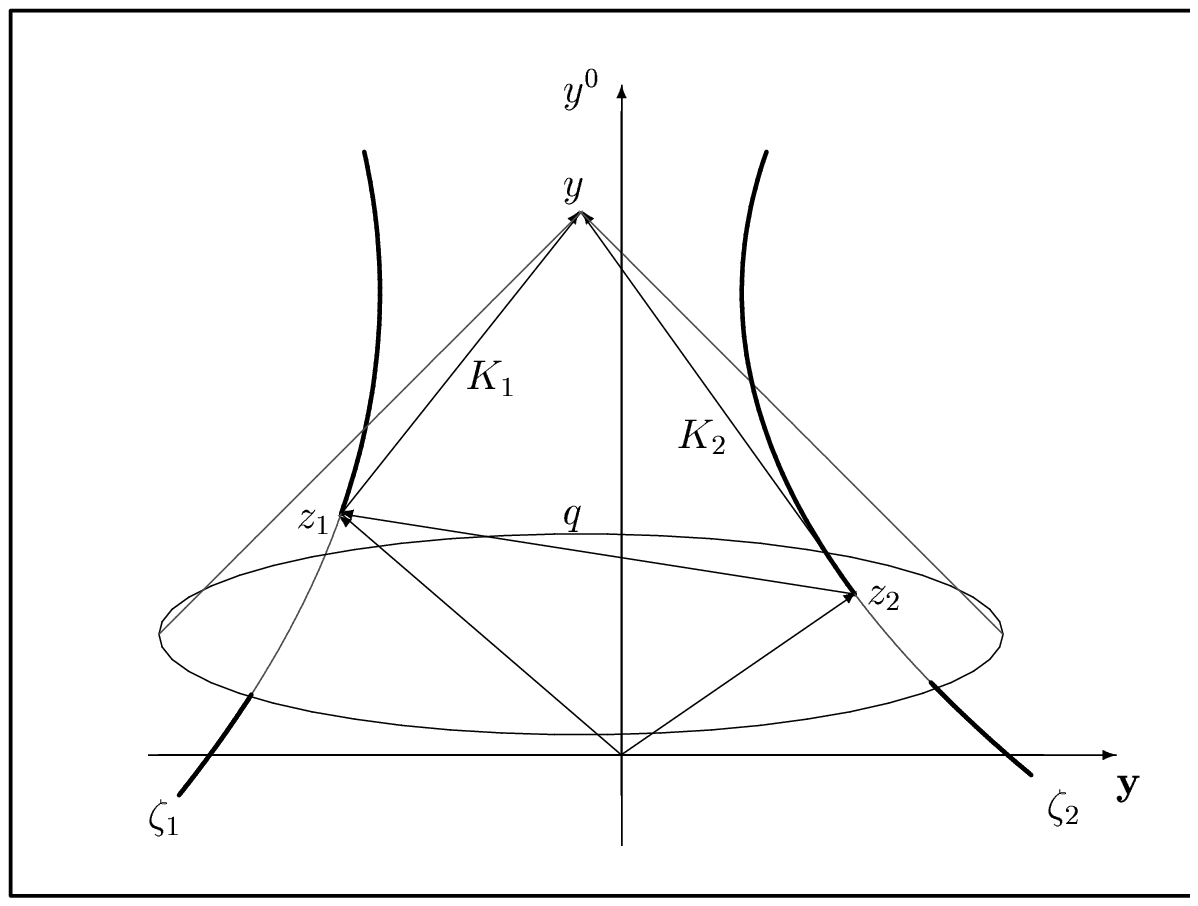,width=0.6\textwidth}
\end{center}
\caption{\label{lc_map}
The past light cone with vertex at point $y\in\Sigma_t$ is punctured by
the world lines of the 1-st particle and the 2-nd particle at points
$z_1(t_1)$ and $z_2(t_2)$, respectively. The vector $K_a$ is a null vector
pointing from the emission point $z_a(t_a)=(t_a,z_a^i(t_a))$ to a field
point $y$. The relative position 4-vector $q=z_1-z_2$ is equal to
difference $K_2-K_1$; its square  $(q\cdot q)=-2(K_2\cdot K_1)$.
}
\end{figure}

To find the Jacobian of coordinate transformation (\ref{Zha}), we
derive its differential chart. Setting $\alpha=0$ in eq.(\ref{Zha})
immediately follows $y^0=t$. Since $t=y^0$, then $\partial
t/\partial y^\alpha=\delta_{0\alpha}$. Because $y$ and $z_a(t_a)$
lie on the light cone (see Fig.~\ref{lc_map}), a change field point
$y$ comes with a change in $t_a$. Suppose that $y$ is displaced to
the new point $y+\delta y$. The new intersection of the past light
cone of this vertex with the $a$-th world line is then
$z_a(t_a+\delta t_a)$. These points are still related by the
equation
\begin{equation}
(y^0+\delta y^0-t_a-\delta t_a)^2=\sum_i(y^i+\delta y^i-z_a^i(t_a+\delta
t_a))^2 .
\end{equation}
Expanding this to the first order in $\delta y$ and $\delta t_a$ and
using the cone equation (\ref{S_a}), we obtain $K_{a,\alpha}\delta
y^\alpha -(v_a\cdot K_a)\delta t_a=0$ or
\begin{eqnarray}\label{dty}
\frac{\displaystyle\partial t_a}{\displaystyle\partial y^\alpha}&=&-
\frac{\displaystyle K_{a,\alpha}}{\displaystyle r_a}\\
&=&-\frac{\displaystyle
\Omega_{\alpha\alpha'}k_a^{\alpha'}}{\displaystyle r_a}.\nonumber
\end{eqnarray}
Here $K_a$ is $a$-th null vector pictured in figure \ref{lc_map} and
symbol $r_a$ denotes the scalar product $(v_a\cdot K_a)$, taken with
opposite sign; noncovariant 4-velocity $v_a:=(1,{\rm d}z_a^i/{\rm
d}t_a)$.

For the angular variable we have
\begin{equation}
\frac{\displaystyle\partial\varphi}{\displaystyle\partial y^\alpha}=
\Omega_{\alpha\alpha'}k_\varphi^{\alpha'},
\end{equation}
where
\begin{equation}
k_\varphi^0=0,\quad
k_\varphi^1=
\frac{\displaystyle\cos\varphi}{\displaystyle h},\quad
k_\varphi^2=-
\frac{\displaystyle\sin\varphi}{\displaystyle h},\quad
k_\varphi^3=0.
\end{equation}
Recall that $h$ is the radius of the circle $C(O,h)=S_1\cap S_2$
pictured in Fig.~\ref{kk}.

Determinant of the matrix which defines this differential chart
gives the inverse Jacobian: $J^{-1}={\rm q}/(r_1r_2)$. The
``interference'' surface element,
\begin{equation}\label{sgm_int}
{\rm d}\sigma_0=\frac{\displaystyle r_1r_2}{\displaystyle \rm q}{\rm
d}t_1{\rm d}t_2{\rm d}\varphi,
\end{equation}
is ill defined if and only if the particles are very close to each
other.

\subsection{Global mapping}

Setting $a=1$ in eq.(\ref{Zha}) we obtain the coordinate system centered
on an accelerated world line of the first particle. The flat spacetime
${\mathbb M}_4$ is a disjoint union of hyperplanes
$\Sigma_t=\{y\in{\mathbb M}_4: y^0=t\}$. An interference hyperplane
$\Sigma_t$ is a disjoint union of retarded spheres $S_1(O_1,t-t_1)$ centered
at points $O_1\in\Sigma_t$ with coordinates ${\mathbf z}_1(t_1)$. A sphere
is covered by its intersections with spherical wave fronts of the
second source. Each circle $S_1\cap S_2$ can be labelled by the
individual time $t_2$ of the second particle and each point on a given
circle can be labelled by its polar angle $\varphi$.

Going along the world line $\zeta_1$ we arrive unavoidably at the
point $t_1^{ret}(t)$, such that the future light cone of
$z_1[t_1^{ret}(t)]$ touches the 2-nd world line at point
$z_2(t)\in\Sigma_t$ (see Fig.~\ref{ncaus}). Light cone of upper
vertices do not intersect the second world line at all.

In context with the principle of retarded causality, $\Sigma_t$ is
divided into two regions where outgoing waves sourced by charged
particles combine in quite different manner.
\begin{itemize}
\item[(i)]
{\it Causal}, which is filled up by spheres $S_1$ of radii larger
than or equal to $t-t_1^{ret}(t)$.
\item[(ii)]
{\it Acausal}, where parameter $t_1$ increases from $t_1^{ret}(t)$
to the instant of observation $t$. It is the ball bounded by sphere
of radius $t-t_1^{ret}(t)$ centered at point ${\mathbf
z}_1[t_1^{ret}(t)]$.
\end{itemize}

\begin{figure}
\begin{center}
\epsfclipon
\epsfig{file=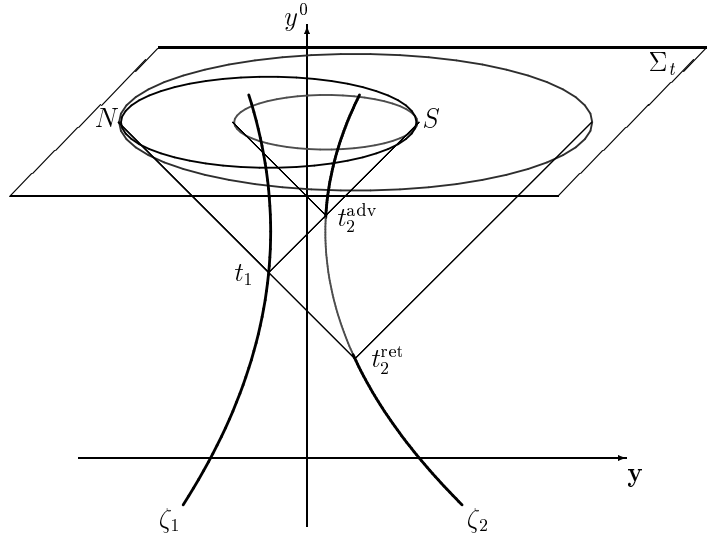,width=0.6\textwidth}
\end{center}
\caption{For a given $t_1$ the wave front $S_1$ is covered by circles
$S_1\cap S_2$ if the parameter $t_2$ increases from $t_2^{\rm ret}(t_1)$ to
$t_2^{\rm adv}(t_1)$. Minimal value labels the vertex of forward light cone
which is punctured by $\zeta_1$ at $z_1(t_1)$. The largest sphere $S_2^{\rm
ret}$ touches the sphere $S_1$ at point $N$. World line $\zeta_2$ punctures
the future light cone of $z_1(t_1)$ at point $z_2(t_2^{\rm adv})$.
Intersection of $S_1$ and the smallest sphere $S_2^{\rm adv}$ contains the
only point $S$.
}\label{ra}
\end{figure}

\subsubsection{Causal region}
Causal region is spanned by curvilinear coordinates (\ref{Zha})
where $t_1$ increases from $-\infty$ to the instant $t_1^{ret}(t)$.
To cover the sphere $S_1$ where $t_1$ is fixed we change the
parameter $t_2$ which labels point $z_2(t_2)\in\zeta_2$. The
starting point is the solution $t_2^{ret}(t_1)$ of algebraic
equation $q^0={\rm q}$ or
\begin{equation}\label{ret2}
t_1-t_2={\rm q}(t_1,t_2),
\end{equation}
where points $z_1(t_1)\in\zeta_1$ and $z_2(t_2^{ret})\in\zeta_2$ are
linked by a null ray. The largest sphere
$S_2(O_2^{ret},t-t_2^{ret})$ touches a given sphere $S_1(O_1,t-t_1)$
at only point N (see Fig.~\ref{ra}). If parameter $t_2$ increases to
$t_2^{adv}(t_1)$ being the solution of algebraic equation $q^0=-{\rm
q}$ or
\begin{equation}\label{adv2}
t_2-t_1={\rm q}(t_1,t_2),
\end{equation}
the intersection $S_1\cap S_2^{adv}$ contains the only point S. If
parameter $t_2$ changes from $t_2^{ret}(t_1)$ to $t_2^{adv}(t_1)$,
the sphere $S_1$ is covered by circles $C(O,h)=S_1\cap S_2$.

\begin{figure}
\begin{center}
\epsfclipon
\epsfig{file=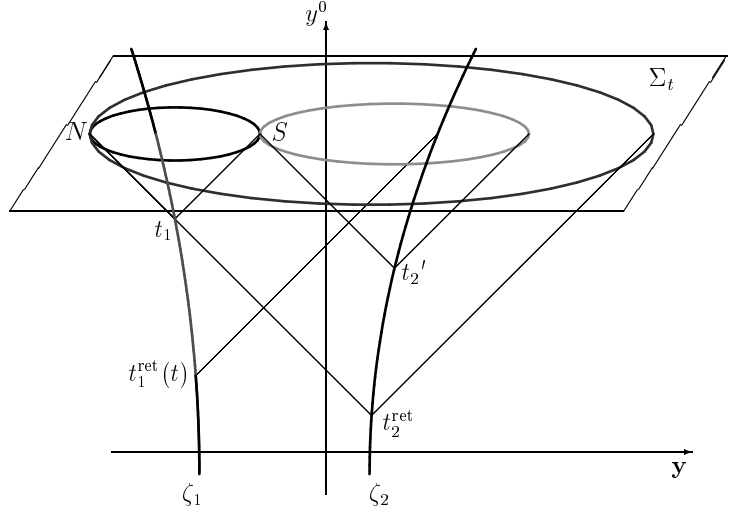,width=0.6\textwidth}
\end{center}
\caption{The instant $t_1^{ret}(t)$ is the solution of algebraic equation
(\ref{adv2}) with $t_2=t$. Sphere
$S_1({\mathbf z}_1^{ret},t-t_1^{ret}(t))$ bounds acausal region.
For a given $t_1\in [t_1^{ret}(t),t]$ the sphere $S_1$ with radius $t-t_1$
is a disjoint union of circles $S_1\cap S_2$ if parameter $t_2$ increases
from $t_2^{ret}(t_1)$ to $t_2'(t,t_1)$. The maximal value satisfies
algebraic equation $k_2^0+k_1^0=|{\bf q}|$. The sphere $S_2{}'$ with radius
$t-t_2'$ touches $S_1$ at point $S$.
Intersection $S_1\cap S_2^{ret}$ contains the only point $N$.
}\label{ncaus}
\end{figure}

\subsubsection{Acausal region}
Acausal region of an interference hyperplane $\Sigma_t$ corresponds
to the fragments of the world lines which are not related to each
other. (By this we mean that the radiation emitted by the first
particle during the interval $[t_1^{ret}(t),t]$ does not come to the
second one and vice versa.) Nevertheless, the outgoing waves of
these portions of world lines combine in $\Sigma_t$ .

Acausal region is filled up by spheres $S_1(O_1,t-t_1)\in\Sigma_t$,
where $t_1\in[t_1^{ret}(t),t]$. The sphere $S_1$ with fixed $t_1$ is
the disjoint union of circles $C(O,h)=S_1\cap S_2$ if the parameter
$t_2$ increases from $t_2^{ret}(t_1)$ to $t_2'(t,t_1)$. The starting
point of this interval is still the solution of eq.(\ref{ret2})
while the maximal value of $t_2$ satisfies the algebraic equation
$k_2^0+k_1^0={\rm q}$ or
\begin{equation}\label{t2prime}
2t-t_1-t_2={\rm q}(t_1,t_2).
\end{equation}
A given sphere $S_1(O_1,t-t_1)$ touches $S_2(O_2',t-t_2')$ at only
point $S$ (see Fig.~\ref{ncaus}).

\subsubsection{Surface integration}
In an analogous way we construct the coordinate system centered on
the world line of the second particle. If $t_2\in
]-\infty,t_2^{ret}(t)]$ then $t_1\in [t_1^{ret}(t_2),
t_1^{adv}(t_2)]$; if $t_2\in [t_2^{ret}(t),t]$ then $t_1\in
[t_1^{ret}(t_2),t_1'(t,t_2)]$, $\varphi \in [0,2\pi [$. The ends of
intervals are defined implicitly by algebraic equations
(\ref{ret2}), (\ref{adv2}), and (\ref{t2prime}).

The surface integration (\ref{p_int}) and (\ref{M_int}) can be
performed via the coordinate system centered on a world line either
of the first particle,
\begin{equation} \label{int1}
\left[\int\limits_{-\infty}^{t_1^{ret}(t)}{\rm d}  t_1
\int\limits_{t_2^{ret}(t_1)}^{t_2^{adv}(t_1)}{\rm d}  t_2 +
\int\limits_{t_1^{ret}(t)}^t{\rm d}  t_1
\int\limits_{t_2^{ret}(t_1)}^{t_2'(t,t_1)}{\rm d}  t_2
\right]\int_0^{2\pi}{\rm d} \varphi\frac{r_1r_2}{\rm q}\,,
\end{equation}
or of the second particle,
\begin{equation} \label{int2}
\left[\int\limits_{-\infty}^{t_2^{ret}(t)}{\rm d}t_2
\int\limits_{t_1^{ret}(t_2)}^{t_1^{adv}(t_2)}{\rm d}  t_1
+ \int\limits_{t_2^{ret}(t)}^t{\rm d}  t_2
\int\limits_{t_1^{ret}(t_2)}^{t_1'(t,t_2)}{\rm d}  t_1
\right]\int_0^{2\pi}{\rm d} \varphi\frac{r_1r_2}{\rm q}\,.
\end{equation}
To calculate the flows (\ref{p_int}) of the mixed electromagnetic field
energy and momentum which flow across the hyperplane $\Sigma_t$, we should
integrate the Maxwell energy-momentum tensor density (\ref{T12}) over
angular variable $\varphi$ and over time variables $t_1$ and $t_2$.

\section{Angular integration of energy-momentum and angular momentum
tensor densities}\label{angul}
\setcounter{equation}{0}
In this Section we trace a series of stages in integration of the mixed
Maxwell energy-momentum tensor density over $\varphi$. In \ref{varphi}
we derive some useful expressions.

In terms of Minkowski coordinates $(y^\alpha)$ the electromagnetic field
generated by $a$-th particle is given by
\begin{equation}\label{f_a}
{\hat f}_{(a)}=\frac{e_a}{{\sf r}_a^2}u_a\wedge {\sf k}_a +
\frac{e_a}{{\sf r}_a}\left[a_a\wedge {\sf k}_a+({\sf k}_a\cdot
a_a)u_a\wedge {\sf k}_a\right],
\end{equation}
where symbol $\wedge$ denotes the wedge product. We use {\sf
sans-serif} symbols for the retarded distance\footnote{Because the
speed of light is set to unity, ${\sf r}_a$ is equal to the spatial
distance between $z_a[s_a^{ret}(y)]$ and $y$ as measured in
momentarily comoving Lorentz frame where $u_a^\alpha=(1,0,0,0)$.},
\begin{equation}\label{rdist}
{\sf r}_a=-\eta_{\alpha\beta}\left(y^\alpha
-z_a^\alpha(s_a)\right)u_a^\beta(s_a),
\end{equation}
and for the null vector $K_a=y-z_a(s_a)$ rescaled by a factor ${\sf
r}_a^{-1}$,
\begin{equation}\label{ksmall}
{\sf k}_a^\alpha=\frac{1}{{\sf r}_a}\left(y^\alpha
-z_a^\alpha(s_a)\right).
\end{equation}
To express field strengths in terms of curvilinear
coordinates $(t,t_1,t_2,\varphi)$, it is advantageous to replace the
retarded proper time $s_a(y)$ by evolution parameter $t_a$. The
components of particles' 4-velocities $u_a$ and 4-accelerations $a_a$,
$a=1,2$, become \cite{Rohr}
\begin{equation}\label{u_a}
u_a^\mu=\gamma_av_a^\mu(t_a),\qquad
a_a^\mu=\gamma_a^4(v_a\cdot{\dot v}_a)v_a^\mu+\gamma_a^2{\dot v}_a^\mu,
\end{equation}
where 4-vectors $v_a^\mu=(1,v_a^i(t_a))$, ${\dot v}_a^\mu=(0,{\dot
v}_a^i(t_a))$ and factor $\gamma_a:=[1-{\mathbf v}_a^2]^{-1/2}$.
Substituting these into eq.(\ref{f_a}) and using the relation
${\sf k}_a^\mu=K_a^\mu/{\sf r}_a$ yields
\begin{equation}\label{f}
{\hat f}_{(a)}=e_a\left(\frac{v_a\wedge K_a}{r_a^3}c_a+
\frac{{\dot v}_a\wedge K_a}{r_a^2}\right),
\end{equation}
where
\begin{equation}\label{cr}
r_a=K_a^0 - ({\bf K}_a{\bf v}_a),\qquad
c_a=\gamma_a^{-2}+({\bf K}_a{\bf\dot v}_a).
\end{equation}
Note that $r_a$ is the retarded distance (\ref{rdist}) rescaled by a
factor $\gamma_a$, i.e. $r_a=\gamma_a^{-1}{\sf r}_a$. The separation
vector $K_a$ has the form ${\hat\Omega}k_a$, where components of
null vector $k_a$ are given by eqs.(\ref{k_k}) and matrix
$\hat\Omega$ determines the transition to momentarily comoving
Lorentz frame associated with basis (\ref{basis}).

It is straightforward to substitute the components of electromagnetic
fields (\ref{f}) in terms of ``interference'' coordinates
$(t,t_1,t_2,\varphi)$ into integrands of expressions (\ref{p_int}) and
(\ref{M_int}) to calculate the interference part of radiated
energy-momentum and angular momentum, respectively. Integration of the
mixed stress-energy tensor over angular variable is the key to the
problem. All $\varphi$-dependent terms are concentrated in the
following constructions:
\begin{eqnarray}\label{ABC}
{\cal D}^a&=&\frac{1}{2\pi}\int\limits_0^{2\pi}{\rm d}\varphi
\frac{a}{{\rm q}r_1r_2},\qquad
{\cal B}^a=\frac{1}{2\pi}\int\limits_0^{2\pi}{\rm d}\varphi
\frac{ac_2}{{\rm q}r_1(r_2)^2},\\
{\cal C}^a&=&\frac{1}{2\pi}\int\limits_0^{2\pi}{\rm d}\varphi
\frac{ac_1}{{\rm q}(r_1)^2r_2},
\qquad
{\cal A}^a=\frac{1}{2\pi}\int\limits_0^{2\pi}{\rm d}\varphi
\frac{ac_1c_2}{{\rm q}(r_1)^2(r_2)^2}.
\nonumber
\end{eqnarray}
They are labeled according to their dependence on the combination of
components of the separation vectors $K_1$ and $K_2$: factor $a$ is
replaced by $K_1^\mu K_2^\nu, K_1^\mu, K_2^\nu$, or $1$ for ${\cal
D}^{\mu\nu}_{12}$, ${\cal D}^\mu_1$, ${\cal D}^\nu_2$ or ${\cal
D}^0$, respectively. (The others ${\cal B}^a, {\cal C}^a$, and
${\cal A}^a$ are marked analogously.)

The mixed part of the stress-energy tensor (\ref{T12}) is symmetric in
indices $1$ and $2$. Substituting (\ref{f}) into the first term of this
expression and using the identities $K_2-K_1=q$ and $(K_2\cdot
K_1)=-1/2(q\cdot q)$ yields
\begin{eqnarray}\label{ff12}
\frac{1}{4\pi}\int\limits_0^{2\pi}{\rm d}\varphi\!\!\!\!\!\!\!\!\!\!\!\!&&
Jf^{\mu\alpha}_{(1)}f^\nu_{(2)\alpha}=\frac{e_1e_2}{2}\left\{
{\cal T}^{\mu\nu}_{12}\left(\frac{\partial^2\sigma}{\partial
t_1\partial t_2}\right)
+{\cal T}^\mu_1\left(v_2^\nu\frac{\partial\sigma}{\partial
t_1}\right)
+{\cal T}^\nu_2\left(v_1^\mu\frac{\partial\sigma}{\partial
t_2}\right)
+{\cal T}^0(v_1^\mu v_2^\nu\sigma)
\right.\nonumber\\
&-&\left. {\cal C}_1^\mu v_2^\nu\frac{\partial^2\sigma}{\partial
t_1\partial t_2}
- {\cal D}_1^\mu v_2^\nu\frac{\partial^3\sigma}{\partial t_1^2\partial t_2}
- {\cal B}_2^\nu v_1^\mu\frac{\partial^2\sigma}{\partial
t_1\partial t_2}
- {\cal D}_2^\nu v_1^\mu\frac{\partial^3\sigma}{\partial t_1\partial t_2^2}
\right.\\
&-&\left.
{\cal B}^0v_1^\mu v_2^\nu\frac{\partial\sigma}{\partial t_1}
-{\cal C}^0v_1^\mu v_2^\nu\frac{\partial\sigma}{\partial t_2}
-{\cal D}^0\left(
{\dot v}_1^\mu v_2^\nu\frac{\partial\sigma}{\partial t_2}
+v_1^\mu {\dot v}_2^\nu\frac{\partial\sigma}{\partial t_1}
+v_1^\mu v_2^\nu\frac{\partial^2\sigma}{\partial t_1\partial t_2}
\right)
\right\},\nonumber
\end{eqnarray}
after addition of similar terms and integration over $\varphi$.
World function $\sigma(t_1,t_2)$ of two spacelike related points,
$z_1(t_1)\in\zeta_1$ and $z_2(t_2)\in\zeta_2$, is equal to one-half
of the square of vector $q=z_1-z_2$, taken with opposite sign,
\begin{equation}\label{Synge}
\sigma(t_1,t_2)=-1/2(q\cdot q).
\end{equation}
Each second order differential operator,
\begin{equation}\label{calTa}
{\hat{\cal T}}^a= {\cal D}^a\frac{\partial^2 }{\partial t_1\partial
t_2}+ {\cal B}^a\frac{\partial}{\partial t_1} +{\cal
C}^a\frac{\partial}{\partial t_2}+{\cal A}^a,
\end{equation}
has been labeled according to dependence of coefficients (\ref{ABC})
on the combination of vectors $K_1$ and $K_2$.

For the convolution $f^{\alpha\beta}_{(1)}f^{(2)}_{\alpha\beta}$, we
obtain
\begin{equation}\label{ff}
\frac{1}{4\pi}\int\limits_0^{2\pi}{\rm d}\varphi
Jf^{\alpha\beta}_{(1)}f^{(2)}_{\alpha\beta}=e_1e_2{\cal T}^0(\lambda),
\end{equation}
where function
\begin{equation}\label{lbd}
\lambda=\sigma\frac{\partial^2\sigma}{\partial t_1\partial t_2} -
\frac{\partial\sigma}{\partial t_1}\frac{\partial\sigma}{\partial
t_2}
\end{equation}
depends on two-point function (\ref{Synge}) and its derivatives in time
variables.

To distinguish the partial derivatives in time variables, we rewrite
the operator (\ref{calTa}) as the sum of the second-order
differential operator,
\begin{equation}\label{Pap}
{\hat\Pi}^a=\frac{\partial^2 }{\partial t_1\partial t_2}{\cal D}^a+
\frac{\partial}{\partial t_1}\left({\cal B}^a- \frac{\partial{\cal
D}^a}{\partial t_2}\right) +\frac{\partial}{\partial t_2}\left({\cal
C}^a- \frac{\partial{\cal D}^a}{\partial t_1}\right),
\end{equation}
and the ``tail'',
\begin{equation}\label{pia}
\pi^a=\frac{\partial^2 {\cal D}^a}{\partial t_1\partial t_2}-
\frac{\partial{\cal B}^a}{\partial t_1}-
\frac{\partial{\cal C}^a}{\partial t_2}+{\cal A}^a.
\end{equation}
For a smooth function $f(t_1,t_2)$ we have
\begin{equation}\label{TaP}
{\hat{\cal T}}^a(f)={\hat\Pi}^a(f)+f\pi^a.
\end{equation}

Cumbersome calculations which are presented in \ref{varphi} give the
relations
\begin{eqnarray}\label{cli}
\pi^0&=&0,\\
\pi_1^\mu&=&v_1^\mu\left({\cal B}^0-\frac{\partial{\cal
D}^0}{\partial t_2}\right),\qquad
\pi_2^\nu=v_2^\nu\left({\cal C}^0-\frac{\partial{\cal D}^0}{\partial
t_1}\right),
\nonumber\\
\pi_{12}^{\mu\nu}&=&
v_1^\mu\left({\cal B}_2^\nu-\frac{\partial {\cal D}_2^\nu}{\partial
t_2}\right)
+v_2^\nu\left({\cal C}_1^\mu-\frac{\partial {\cal
D}_1^\mu}{\partial t_1}\right)-v_1^\mu v_2^\nu {\cal D}^0,
\nonumber
\end{eqnarray}
which allow us to rewrite the sum of integrals (\ref{ff12}) and
(\ref{ff}) in terms of differential operators ${\hat\Pi}^a$ and
partial derivatives in $t_1$ and $t_2$,
\begin{eqnarray}\label{f12fi}
{\cal P}^{\mu\nu}_{12}&=&\frac{1}{4\pi}
\int_0^{2\pi}d\varphi J\left(f_{(1)}^{\mu\alpha}f_{(2)\alpha}^\nu -
\frac{\eta^{\mu\nu}}{4}f_{(1)}^{\alpha\beta}f_{\alpha\beta}^{(2)}
\right)\\
&=&\frac{e_1e_2}{2}\left\{
{\hat\Pi}^{\mu\nu}_{12}\left(\frac{\partial^2\sigma}{\partial
t_1\partial t_2}\right)
+{\hat\Pi}^\mu_1\left(v_2^\nu\frac{\partial\sigma}{\partial
t_1}\right)
+{\hat\Pi}^\nu_2\left(v_1^\mu\frac{\partial\sigma}{\partial
t_2}\right)
+{\hat\Pi}^0(v_1^\mu v_2^\nu\sigma)
\right.\nonumber\\
&-&\left.
\frac{\partial}{\partial t_1}\left(
{\cal D}_1^\mu v_2^\nu\frac{\partial^2\sigma}{\partial t_1\partial t_2}
\right)
-\frac{\partial}{\partial t_2}\left(
{\cal D}_2^\nu v_1^\mu\frac{\partial^2\sigma}{\partial t_1\partial t_2}
\right)
\right.\nonumber\\
&-&\left.
\frac{\partial}{\partial t_1}\left(
{\cal D}^0v_1^\mu v_2^\nu\frac{\partial\sigma}{\partial t_2}
\right)
-\frac{\partial}{\partial t_2}\left(
{\cal D}^0 v_1^\mu v_2^\nu\frac{\partial\sigma}{\partial t_1}
\right)-\frac{1}{2}\eta^{\mu\nu}
{\hat\Pi}^0(\lambda)
\right\}.\nonumber
\end{eqnarray}
We denote ${\cal P}^{\mu\nu}_{21}$ the integral over $\varphi$ of the
remaining terms involved in tensor (\ref{T12}). It can be obtained by
interchanging of indices 1 and 2.

Setting $\mu=0$ and $\nu=i$ in eq.(\ref{f12fi}), we obtain the first
term of the mixed space-time components of the stress-energy tensor
({\ref{T12}). We add the term where indices $1$ and $2$ are
interchanged. Since zeroth components $k_1^0$ and $k_2^0$ of the
separation four-vectors $K_1$ and $K_2$ do not depend on $\varphi$,
the final expression get simplified,
\begin{eqnarray} \label{t0if}
{\cal P}^i_{\rm int}&=&{\cal P}^{0i}_{12}+{\cal P}^{0i}_{21}\\
&=&\frac{e_1e_2}{2}\left[
{\hat\Pi }_2^i\left(\frac{\partial\lambda_1}{\partial t_2}\right)+
{\hat\Pi }_1^i\left(\frac{\partial\lambda_2}{\partial t_1}\right) +
{\hat\Pi }^0\left(v_2^i\lambda_1+v_1^i\lambda_2\right)\right.\nonumber\\
&-&\left.\frac{\partial}{\partial t_1}\left(v_2^i
\frac{\partial\lambda_1}{\partial t_2}{\cal D}^0
\right)-
\frac{\partial}{\partial t_2}\left(v_1^i
\frac{\partial\lambda_2}{\partial t_1}{\cal D}^0
\right)
\right],\nonumber
\end{eqnarray}
where
\begin{equation}\label{l1-2}
\lambda_1=k_1^0\frac{\partial\sigma}{\partial t_1}+\sigma,\qquad
\lambda_2=k_2^0\frac{\partial\sigma}{\partial t_2}+\sigma.
\end{equation}

Similarly we derive zeroth component ${\cal P}^0_{\rm int}$. Setting
$\mu=0$ and $\nu=0$ in eq.(\ref{f12fi}), we obtain the first
one-half of desired expression.  The second one, ${\cal
P}^{00}_{21}$, can be derived via interchanging indices $1$ and $2$.
The integral of energy density $T^{00}_{\rm int}$ over the angular
variable has the form
\begin{eqnarray} \label{t00f}
{\cal P}^0_{\rm int}&=&{\cal P}^{00}_{12}+{\cal P}^{00}_{21}\\
&=&\frac{e_1e_2}{2}{\hat\Pi
}^0\left(\Sigma\frac{\partial^2\Sigma}{\partial t_1\partial t_2} -
\frac{\partial\Sigma}{\partial t_1}\frac{\partial\Sigma}{\partial
t_2}\right), \nonumber
\end{eqnarray}
where three-point function,
\begin{equation}\label{Sigm}
\Sigma(t,t_1,t_2)=2k_1^0k_2^0 +\sigma(t_1,t_2),
\end{equation}
depends on particles' positions referred to the moments $t_1$ and $t_2$
before observation instant $t$ as well as on $t$ itself.

We now turn to the integration of the angular momentum tensor
density (\ref{M_int}) carried by the electromagnetic field due to two
pointlike charges. We present the torque $m_{\rm int}^{\mu\nu}=y^\mu T_{\rm
int}^{0\nu} - y^\nu T_{\rm int}^{0\mu}$ in the following form:
\begin{equation}\label{mm}
m_{\rm int}^{\mu\nu}=
m_{12}^{\mu\nu}+m_{21}^{\mu\nu}-m_{12}^{\nu\mu}-m_{21}^{\nu\mu},
\end{equation}
where
\begin{equation}\label{m12}
m_{12}^{\mu\nu}=\left(z_1^\mu+K_1^\mu\right)\frac{1}{4\pi}
\left[
f_{(1)}^{0\lambda}f_{(2)\lambda}^\nu -\frac14\eta^{0\nu}
f_{(1)}^{\alpha\beta}f^{(2)}_{\alpha\beta}
\right].
\end{equation}
It is straightforward to substitute the fields (\ref{f}) into this
expression to calculate the first term of expression (\ref{mm}). The others
can be obtained by interchanging of the pair of indices $(1,2)$ and
$(\mu,\nu)$.

Having integrated expression $Jm_{12}^{\mu\nu}$ over $\varphi$ we
obtain
\begin{eqnarray}\label{m-12}
{\cal M}_{12}^{\mu\nu}&=&\frac{e_1e_2}{2}\left\{
{\hat{\cal T}}_{12}^{\mu\nu}\left(\frac{\partial\lambda_1}{\partial
t_2}\right)+{\hat{\cal T}}_1^\mu\left(v_2^\nu\lambda_1\right)-
v_2^\nu\frac{\partial\lambda_1}{\partial t_2}{\cal C}_1^\mu-
v_2^\nu\frac{\partial^2\lambda_1}{\partial t_1\partial t_2}{\cal D}_1^\mu
\right.\nonumber\\
&+&\left.
{\hat{\cal T}}_2^\nu\left(z_1^\mu\frac{\partial\lambda_1}{\partial t_2}
\right) -
v_1^\mu\frac{\partial\lambda_1}{\partial t_2}{\cal B}_2^\nu-
v_1^\mu\frac{\partial^2\lambda_1}{\partial t_2^2}{\cal D}_2^\nu
+{\hat{\cal T}}^0\left(z_1^\mu v_2^\nu\lambda_1\right)
-v_1^\mu v_2^\nu\lambda_1{\cal B}^0\right.\nonumber\\
&-&\left.
v_1^\mu\left({\dot v}_2^\nu\lambda_1+v_2^\nu
\frac{\partial\lambda_1}{\partial t_2}{\cal D}^0\right)
-z_1^\mu v_2^\nu\frac{\partial\lambda_1}{\partial t_2}{\cal C}^0-
z_1^\mu v_2^\nu\frac{\partial^2\lambda_1}{\partial t_1\partial t_2}{\cal
D}^0
\right.\nonumber\\
&-&\left.
\frac{\eta^{0\nu}}{2}\left[
{\hat{\cal T}}_1^\mu\left(\lambda\right)+
{\hat{\cal T}}^0\left(z_1^\mu\lambda\right)
-v_1^\mu\lambda{\cal B}^0 -v_1^\mu\frac{\partial\lambda}{\partial t_2}{\cal
D}^0\right]
\right\},
\end{eqnarray}
where functions $\lambda$ and $\lambda_a$ are given by eqs.(\ref{lbd}) and
(\ref{l1-2}), respectively.

Usage of the equalities in eq.(\ref{cli}) derived in \ref{varphi} allows us
to rewrite the integrand (\ref{m-12}) as follows:
\begin{eqnarray}\label{M-12}
{\cal M}_{12}^{\mu\nu}&=&\frac{e_1e_2}{2}\left\{
{\hat\Pi}_{12}^{\mu\nu}\left(\frac{\partial\lambda_1}{\partial
t_2}\right)+{\hat\Pi}_1^\mu\left(v_2^\nu\lambda_1\right)-
\frac{\partial}{\partial t_1}\left(v_2^\nu\frac{\partial\lambda_1}{\partial
t_2}{\cal D}_1^\mu\right)
\right.\\
&+&\left.{\hat\Pi}_2^\nu\left(z_1^\mu\frac{\partial\lambda_1}{\partial
t_2}\right)-
\frac{\partial}{\partial t_2}\left(v_1^\mu\frac{\partial\lambda_1}{\partial
t_2}{\cal D}_2^\nu\right)
\right.\nonumber\\
&+&\left.{\hat\Pi}^0\left(z_1^\mu v_2^\nu\lambda_1\right)-
\frac{\partial}{\partial
t_1}\left(z_1^\mu v_2^\nu\frac{\partial\lambda_1}{\partial t_2}{\cal
D}^0\right)-
\frac{\partial}{\partial t_2}\left(v_1^\mu
v_2^\nu\lambda_1{\cal D}^0\right)
\right.\nonumber\\
&-&\left.\frac{\eta^{0\nu}}{2}\left[
{\hat\Pi}_1^\mu\left(\lambda\right)+
{\hat\Pi}^0\left(z_1^\mu\lambda\right)-
\frac{\partial}{\partial t_2}\left(v_1^\mu\lambda{\cal D}^0\right)
\right]\right\}.\nonumber
\end{eqnarray}
The other terms of mixed angular momentum,
\begin{equation}\label{MM}
{\cal M}_{\rm int}^{\mu\nu}= {\cal M}_{12}^{\mu\nu}+{\cal
M}_{21}^{\mu\nu}-{\cal M}_{12}^{\nu\mu}-{\cal M}_{21}^{\nu\mu},
\end{equation}
can be obtained via interchanging of indices $(1,2)$ and $(\mu,\nu)$.

We see that the integration of the mixed stress-energy tensor (\ref{T12})
over $\varphi$ yields the combinations of partial derivatives in time
variables. In the next Section we classify them and reveal the
long-range terms which contribute into radiated energy-momentum.

\section{Radiative parts of mixed energy-momentum and angular
momentum}\label{mix}
\setcounter{equation}{0}

In previous Section we integrate the mixed part of the stress-energy
tensor and its torque over polar angle. Resulted expressions
describe contributions to electromagnetic field's energy-momentum
and angular momentum due to interference of spherical wave fronts of
charges $e_1$ and $e_2$ placed at fixed points $z_1(t_1)\in\zeta_1$
and $z_2(t_2)\in\zeta_2$, respectively. The crucial issue is that
the integrals (\ref{t0if}), (\ref{t00f}) and (\ref{MM}) have the
remarkable property of being the sum of partial derivatives in time
variables. This circumstance allows us to calculate how much
electromagnetic field's energy-momentum and angular momentum flow
across a hyperplane $\Sigma_t$.

It is natural to integrate the expression being the time derivative
with respect to $t_2$ according to the rule (\ref{int1}),
\begin{eqnarray} \label{d2G}
{\cal G}_2&=&\left[\int\limits_{-\infty}^{t_1^{ret}(t)}{\rm d}  t_1
\int\limits_{t_2^{ret}(t_1)}^{t_2^{adv}(t_1)}{\rm d}  t_2
+ \int\limits_{t_1^{ret}(t)}^t{\rm d}  t_1
\int\limits_{t_2^{ret}(t_1)}^{t_2'(t,t_1)}{\rm d}  t_2
\right]\frac{\partial G_2(t_1,t_2)}{\partial t_2}\\
&=&\int\limits_{-\infty}^{t_1^{ret}(t)}{\rm d}  t_1G_2[t_1,t_2^{adv}(t_1)] -
\int\limits_{-\infty}^{t}{\rm d} t_1G_2[t_1,t_2^{ret}(t_1)] +
\int\limits_{t_1^{ret}(t)}^t{\rm d}t_1G_2[t_1,t_2'(t,t_1)].\nonumber
\end{eqnarray}
Having applied the rule (\ref{int2}) to the expression of type $\partial
G_1/\partial t_1$, we obtain
\begin{eqnarray} \label{d1G}
{\cal G}_1&=&\left[\int\limits_{-\infty}^{t_2^{ret}(t)}{\rm d}t_2
\int\limits_{t_1^{ret}(t_2)}^{t_1^{adv}(t_2)}{\rm d}t_1
+ \int\limits_{t_2^{ret}(t)}^t{\rm d}t_2
\int\limits_{t_1^{ret}(t_2)}^{t_1'(t,t_2)}{\rm d}  t_1
\right]\frac{\partial G_1(t_1,t_2)}{\partial t_1}\\
&=&\int\limits_{-\infty}^{t_2^{ret}(t)}{\rm d} t_2 G_1[t_1^{adv}(t_2),t_2]
-\int\limits_{-\infty}^{t}{\rm d} t_2 G_1[t_1^{ret}(t_2),t_2]
+\int\limits_{t_2^{ret}(t)}^t{\rm d}t_2 G_1[t_1'(t,t_2),t_2].\nonumber
\end{eqnarray}
The end points are valuable only in the integration procedure. They
are solutions of algebraic equations (\ref{ret2}), (\ref{adv2}) and
(\ref{t2prime}). The retarded instants $t_a^{ret}(t_b)$ and advanced
ones $t_b^{adv}(t_a)$ label the points $N$ and $S$ in which fronts
of {\it outgoing} electromagnetic waves produced by charges touch
each other (see Fig.~\ref{ra}). All the moments are {\it before} the
observation instant $t$, so that the retarded causality is not
violated.

It is worth noting that the functions $t_1^{ret}(t_2)$ and $t_2^{adv}(t_1)$
are inverted to each other as well as the pair of functions $t_1^{adv}(t_2)$
and $t_2^{ret}(t_1)$. For a fixed laboratory time $t$ the functions
$t_1'(t,t_2)$  and $t_2'(t,t_1)$ are inverses too. These circumstances allow
us to change the variables $t_a\mapsto t_a^{ret}(t_b)$ in the ``advanced''
integrals in eqs.(\ref{d2G}) and (\ref{d1G}). Further we couple them
with their ``retarded'' counterparts. Since
\begin{equation}
\frac{{\rm d}t_1^{ret}(t_2)}{{\rm d}t_2}=\frac{1-({\mathbf v}_1{\mathbf
n}_{\rm q})}{1-({\mathbf v}_2{\mathbf n}_{\rm q})},\qquad
\frac{{\rm d}t_2^{ret}(t_1)}{{\rm d}t_1}=\frac{1+({\mathbf v}_2{\mathbf
n}_{\rm q})}{1+({\mathbf v}_1{\mathbf n}_{\rm q})},
\end{equation}
we obtain
\begin{eqnarray} \label{d2Gd1G}
{\cal G}_2+{\cal G}_1&=&\int\limits_{-\infty}^t{\rm d}t_1
\left[
\frac{1-({\mathbf v}_1{\mathbf n}_{\rm q})}{1-({\mathbf v}_2{\mathbf
n}_{\rm q})}G_1-G_2
\right]_{t_2=t_2^{ret}(t_1)}\\
&+&
\int\limits_{-\infty}^t{\rm d}t_2
\left[-G_1+
\frac{1+({\mathbf v}_2{\mathbf n}_{\rm q})}{1+({\mathbf v}_1{\mathbf
n}_{\rm q})}G_2
\right]_{t_1=t_1^{ret}(t_2)}\nonumber\\
&+&
\int\limits_{t_2^{ret}(t)}^t{\rm d}t_2
\left[G_1+
\frac{1-({\mathbf v}_2{\mathbf n}_{\rm q})}{1+({\mathbf v}_1{\mathbf
n}_{\rm q})}G_2
\right]_{t_1=t_1'(t,t_2)}.\nonumber
\end{eqnarray}
The unit vector ${\mathbf n}_q={\mathbf q}/{\rm q}$ is the third
vector of orthonormal triad (\ref{basis}). The last integral is due
to interference of outgoing electromagnetic waves radiated out by
the particles within the acausal region (see Fig.~\ref{ncaus}). We
take into account that
\begin{equation}
\frac{{\rm d}t_2'(t,t_1)}{{\rm d}t_1}=-\frac{1+({\mathbf v}_1{\mathbf
n}_{\rm q})}{1-({\mathbf v}_2{\mathbf n}_{\rm q})}.
\end{equation}

Of course, one can change the variables $t_b\mapsto t_b^{adv}(t_a)$
in the ``retarded'' integrals in eqs.(\ref{d2G}) and (\ref{d1G}) and
add them to their ``advanced'' counterparts,
\begin{eqnarray} \label{d1Gd2G}
{\cal G}_2+{\cal G}_1&=&\int\limits_{-\infty}^{t_2^{ret}(t)}{\rm d}t_2
\left[
G_1-\frac{1-({\mathbf v}_2{\mathbf n}_{\rm q})}{1-({\mathbf v}_1{\mathbf
n}_{\rm q})}G_2
\right]^{t_1=t_1^{adv}(t_2)}\\
&+&
\int\limits_{-\infty}^{t_1^{ret}(t)}{\rm d}t_1
\left[-\frac{1+({\mathbf v}_1{\mathbf n}_{\rm q})}{1+({\mathbf v}_2{\mathbf
n}_{\rm q})}G_1+G_2
\right]^{t_2=t_2^{adv}(t_1)}\nonumber\\
&+&
\int\limits_{t_1^{ret}(t)}^t{\rm d}t_1
\left[\frac{1+({\mathbf v}_1{\mathbf n}_{\rm q})}{1-({\mathbf v}_2{\mathbf
n}_{\rm q})}G_1+G_2
\right]^{t_2=t_2'(t,t_1)}.\nonumber
\end{eqnarray}
A combination of the ``retarded'' and the ``advanced'' terms is valuable
too.

Integral of a mixed double derivative can be written in the form either
\begin{eqnarray} \label{d1G0}
{\cal I}_1&=&\left[\int\limits_{-\infty}^{t_2^{ret}(t)}{\rm d}t_2
\int\limits_{t_1^{ret}(t_2)}^{t_1^{adv}(t_2)}{\rm d}t_1
+ \int\limits_{t_2^{ret}(t)}^t{\rm d}t_2
\int\limits_{t_1^{ret}(t_2)}^{t_1'(t,t_2)}{\rm d}t_1
\right]\frac{\partial}{\partial t_1}\left[
\frac{\partial G(t_1,t_2)}{\partial t_2}
\right]\\
&=&\int\limits_{-\infty}^{t_2^{ret}(t)}{\rm d}t_2
\left[\frac{\partial G(t_1,t_2)}{\partial
t_2}\right]^{t_1=t_1^{adv}(t_2)}
-\int\limits_{-\infty}^{t}{\rm d}t_2
\left[\frac{\partial G(t_1,t_2)}{\partial
t_2}\right]_{t_1=t_1^{ret}(t_2)}\nonumber\\
&+&\int\limits_{t_2^{ret}(t)}^t{\rm d}t_2
\left[\frac{\partial G(t_1,t_2)}{\partial
t_2}\right]^{t_1=t_1'(t,t_2)},\nonumber
\end{eqnarray}
or
\begin{eqnarray} \label{d2G0}
{\cal I}_2&=&\left[\int\limits_{-\infty}^{t_1^{ret}(t)}{\rm d}t_1
\int\limits_{t_2^{ret}(t_1)}^{t_2^{adv}(t_1)}{\rm d}t_2
+ \int\limits_{t_1^{ret}(t)}^t{\rm d}t_1
\int\limits_{t_2^{ret}(t_1)}^{t_2'(t,t_1)}{\rm d}t_2
\right]\frac{\partial}{\partial t_2}\left[\frac{\partial
G(t_1,t_2)}{\partial t_1}\right]\\
&=&\int\limits_{-\infty}^{t_1^{ret}(t)}{\rm d}t_1
\left[\frac{\partial G(t_1,t_2)}{\partial
t_1}\right]^{t_2=t_2^{adv}(t_1)} -
\int\limits_{-\infty}^{t}{\rm d}t_1
\left[\frac{\partial G(t_1,t_2)}{\partial
t_1}\right]_{t_2=t_2^{ret}(t_1)}\nonumber\\
&+&
\int\limits_{t_1^{ret}(t)}^t{\rm d}t_1
\left[\frac{\partial G(t_1,t_2)}{\partial
t_1}\right]^{t_2=t_2'(t,t_1)}.\nonumber
\end{eqnarray}
The question is what expression should be used.

\begin{figure}
\begin{center}
\epsfclipon
\epsfig{file=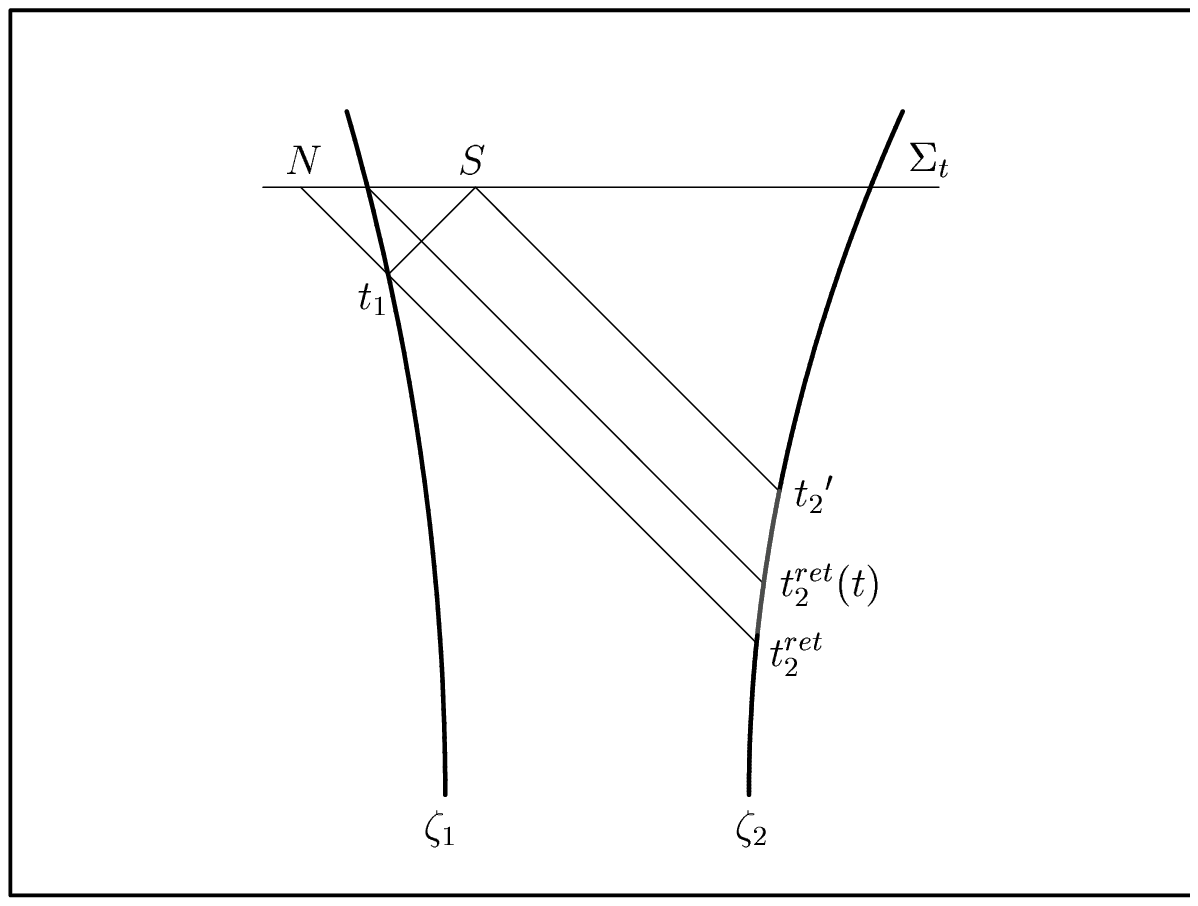,width=0.4\textwidth}
\end{center}
\caption{Function $G[t_1,t_2^{ret}(t_1)]$ is associated with
the interference of wave fronts $S_1(O_1,t-t_1)$ and
$S_2(O_2^{ret},t-t_2^{ret})$ at point $N\in\Sigma_t$. Function
$G[t_1,t_2'(t,t_1)]$ is connected with the combination of wave fronts
$S_1(O_1,t-t_1)$ and $S_2(O_2',t-t_2')$ at point $S\in\Sigma_t$. If $t_1\to
t$, both the point $N$ and the point $S$ tend to
$z_1(t)=\zeta_1\cap\Sigma_t$. If $\lim\limits_{t_1\to
t}G[t_1,t_2^{ret}(t_1)]=\lim\limits_{t_1\to t}G[t_1,t_2'(t,t_1)]$ the value
of integral of mixed derivative $\partial^2G/\partial t_1\partial t_2$ over
time variables does not depend on the order of integration.
}\label{ret-pr}
\end{figure}

To compare ${\cal I}_1$ and ${\cal I}_2$ we change the variables in
the ``advanced'' integrals and subtract (\ref{d2G0}) and
(\ref{d1G0}). We arrive at the integrals being functions of the end
points only,
\begin{eqnarray}\label{deltaI}
{\cal I}_1-{\cal
I}_2&=&\left.\phantom{\frac11}\!\!\!\!G[t_1,t_2^{ret}(t_1)]\right|_{t_1\to
-\infty}^{t_1\to t} -
\left.\phantom{\frac11}\!\!\!\!G[t_1^{ret}(t_2),t_2]\right|_{t_2\to
-\infty}^{t_2\to t}\\
&+&\left\{
\begin{array}{cc}
{\rm either}&
\left.\phantom{\frac{\displaystyle
1}{\displaystyle 1}}\!\!\!\!G[t_1'(t,t_2),t_2]\right|_{t_2\to
t_2^{ret}(t)}^{t_2\to t}\\[1em]
{\rm or} &
-\left.\phantom{\frac{\displaystyle
1}{\displaystyle 1}}\!\!\!\!G[t_1,t_2'(t,t_1)]\right|_{t_1\to
t_1^{ret}(t)}^{t_1\to t}
\end{array}
\right..\nonumber
\end{eqnarray}
It vanishes if and only if (i) limiting values of $G$ (that
evaluated at the remote past) cancel each other, and (ii) function
$G$ is smooth at points at which the world lines puncture
$\Sigma_t$. Indeed, the part of difference (\ref{deltaI}) which
depends on the momentary state of particles' motion can be rewritten
as follows:
\begin{eqnarray}
{\cal I}_1-{\cal I}_2&=&\lim_{t_1\to t}\left\{
G[t_1,t_2^{ret}(t_1)]-G[t_1,t_2'(t,t_1)]
\right\}\\
&-&\lim_{t_2\to t}\left\{
G[t_1^{ret}(t_2),t_2]-G[t_1'(t,t_2),t_2]
\right\}.\nonumber
\end{eqnarray}
The situation is illustrated in Fig.~\ref{ret-pr}.

\subsection{Criteria}

The main task of the present paper is to decompose the interference part of
the Maxwell energy-momentum tensor density into bound and radiative
components. The former modifies individual 4-momenta (\ref{p_a}) of
dressed particles while the latter shows how a charge is influenced by
radiation of another charge.

To reveal meaningful radiative part of mixed energy-momentum
(\ref{p_int}) and angular momentum (\ref{M_int}), we apply the
criteria which were first formulated in Ref.~\cite[Table 1]{Teit}.
\begin{itemize}
\item
The bound part diverges while the radiative one is finite.
\item
The bound component depends on the momentary state of the particles'
motion while the radiative one is accumulated with time.
\item
The form of the bound terms heavily depends on choosing of an
integration surface while the radiative terms are invariant.
\end{itemize}
There are, however, a several properties that the desired expressions have
possess before they can be accepted. We list them below.
\begin{enumerate}
\item
Radiative parts should be completely determined by particles' motion;
they can not depend on distance to a point of observation.
\item
They should be produced by divergence-free expressions.
\item
Non-accelerated charges do not radiate.
\item
Balance of Noether conserved quantities yields the Lorentz-Dirac
equation.
\end{enumerate}

\subsection{Radiative part of mixed momentum}
To decompose the momentum (\ref{t0if}) into bound and radiative
components is a straightforward integration of all the terms over
time variables. Scrupulous computations reveal candidate for
radiative part,
\begin{eqnarray} \label{Pirad}
{\cal P}^i_{\rm int, rad}&=&\frac{e_1e_2}{2}\left[
{\hat\Pi }_2^i\left(k_1^0\frac{\partial^2\sigma}{\partial t_1\partial
t_2}\right)
-\frac{\partial}{\partial t_1}\left(
v_2^ik_1^0\frac{\partial^2\sigma}{\partial t_1\partial t_2}{\cal D}^0
\right)\right.\\
&+&\left.{\hat\Pi }_1^i\left(k_2^0\frac{\partial^2\sigma}{\partial
t_1\partial t_2}\right) -\frac{\partial}{\partial t_2}\left(
v_1^ik_2^0\frac{\partial^2\sigma}{\partial t_1\partial t_2}{\cal
D}^0 \right)\right],\nonumber
\end{eqnarray}
which produce the terms that satisfy Teitelboim's criteria.

Equations (\ref{d2G}) and (\ref{d1G}) imply that its integral over
time variables is completely determined by values of the arguments
of time differential operators at the ends of integration intervals.
When a consideration is restricted to the end points where radius
$h$ of intersection $S_1\cap S_2$ vanishes, the coefficients
(\ref{ABC}) get simplified. Since eq. (\ref{k_k}) the integrands do
not depend on the polar angle at all. The integral over $\varphi$
becomes unit operator. Radiative momentum (\ref{Pirad}) contains
also non-trivial constructions,
\begin{equation}
{\cal C}^a-\frac{\partial{\cal D}^a}{\partial t_1},\qquad {\cal
B}^a-\frac{\partial{\cal D}^a}{\partial t_2},
\end{equation}
which are calculated in \ref{varphi}. All the expressions
(\ref{D0}), (\ref{DBC1i}), (\ref{DBC2i}), and (\ref{DBC12}) are
regular if $h=0$. Values of relative position 4-vector $q$, retarded
distances $r_a$, and some basic functions at limiting points are
collected in Table \ref{krsl} (see \ref{bound_i}).

The sum of differential operators ${\hat\Pi }_1^i$ and ${\hat\Pi
}_2^i$ contains mixed double derivative of the function,
\begin{equation}\label{ddi}
G^i=k_1^0\frac{\partial^2\sigma}{\partial t_1\partial t_2}{\cal D}_2^i
+k_2^0\frac{\partial^2\sigma}{\partial t_1\partial t_2}{\cal D}_1^i.
\end{equation}
A surprising feature of integration of expression (\ref{Pirad}) over time variables
is that the result heavily depends on the order of differentiation in
$\partial^2G^i/\partial t_1\partial t_2$. If one choose the rule
(\ref{d1G0}) they obtain
\begin{equation}\label{p1rad}
p_{{\rm rad},21}^i=
\int_{-\infty}^t{\rm d}t_1\gamma_1^{-1}G_{21}^i[t_1,t_2^{ret}(t_1)]
+
\int_{-\infty}^{t_1^{ret}(t)}{\rm
d}t_1\gamma_1^{-1}G_{21}^i[t_1,t_2^{adv}(t_1)].
\end{equation}
Having applied the rule (\ref{d2G0}) we arrive at
\begin{equation}\label{p2rad}
p_{{\rm rad},12}^i=
\int_{-\infty}^t{\rm d}t_2\gamma_2^{-1}G_{12}^i[t_1^{ret}(t_2),t_2]
+
\int_{-\infty}^{t_2^{ret}(t)}{\rm
d}t_2\gamma_2^{-1}G_{12}^i[t_1^{adv}(t_2),t_2].
\end{equation}
Function under integral signs,
\begin{eqnarray}\label{gba}
\gamma_a^{-1}G_{ba}^i&=&e_ae_b\left[\frac{q_{ab}^i(v_a\cdot
v_b)}{r_b^3}\gamma_b^{-2}
+\frac{q_{ab}^i(v_a\cdot v_b)}{r_b^3}(q_{ab}\cdot{\dot v}_b)\right.\\
&+&\left.\frac{q_{ab}^i(v_a\cdot {\dot v}_b)}{r_b^2}-
\frac{v_b^i(v_a\cdot v_b)}{r_b^2}\right],\nonumber
\end{eqnarray}
is referred to the retarded and the advanced instants. Relative
distance 4-vector $q_{ab}:=z_a-z_b$; in our notation $q_{12}=q$
while $q_{21}=-q$. Denominator $r_b=-(q_{ab}\cdot v_b)$ is the
retarded distance of type (\ref{cr}) where both the field point and
the point of emission are placed on particles' world lines. So, in
expression (\ref{p1rad}) the field point is $z_1(t_1)\in\zeta_1$
where $e_1$ is placed. Points of emission are
$z_2[t_2^{ret}(t_1)]\in\zeta_2$ in the first path integral and
$z_2[t_2^{adv}(t_1)]\in\zeta_2$ in the second one,
\begin{equation}
\left.\phantom{\int}r_2^{ret}={\rm q}\left[1-({\mathbf v}_2{\mathbf n}_{\rm
q})\right]\right|_{[t_1,t_2^{ret}(t_1)]},\qquad
\left.\phantom{\int}r_2^{adv}=-{\rm q}\left[1+({\mathbf v}_2{\mathbf n}_{\rm
q})\right]\right|_{[t_1,t_2^{adv}(t_1)]}.
\end{equation}
In expression (\ref{p2rad}) the field point is in $z_2(t_2)\in\zeta_2$ where
the second charge is placed. It is acted on by the first charge placed at
points either $z_1[t_1^{ret}(t_2)]\in\zeta_1$ or
$z_1[t_1^{adv}(t_2)]\in\zeta_1$. The distances are as follows:
\begin{equation}
\left.\phantom{\int}r_1^{ret}={\rm q}\left[1+({\mathbf v}_1{\mathbf
n}_{\rm q})\right]\right|_{[t_1^{ret}(t_2),t_2]},\qquad
\left.\phantom{\int}r_1^{adv}=-{\rm q}\left[1-({\mathbf v}_1{\mathbf
n}_{\rm q})\right]\right|_{[t_1^{adv}(t_2),t_2]}.
\end{equation}

It is of great importance that the integrand (\ref{gba}) does not
depend on the distances to points $N$ or $S$ on the observation
hyperplane $\Sigma_t$. It depends on the relative distance between
charged particles, on their velocities, and on their accelerations.
The situation looks like $a$-th charge is acted on by $b$-th one
directly. The charges are connected by a null ray: in the first
integral in eq.(\ref{p1rad}) or in eq.(\ref{p2rad}) the interaction
be forward while in the second one backward in time. All the moments
are before the instant of observation $t$ which labels the upper
limits of path integrals.

Since the function (\ref{ddi}) is not smooth in neighborhoods of
intersections $\zeta_a\cap\Sigma_t$, the expressions (\ref{p1rad}) and
eq.(\ref{p2rad}) are not equivalent. By virtue of the relation
(\ref{deltaI}) we compute the difference:
\begin{equation}\label{Deltai}
p_{{\rm rad},21}^i-p_{{\rm rad},12}^i=
-\left.\left.q^i(A_{21}^{ret}\cdot
A_{12}^{adv})\right|_{[t_1,t_2^{ret}(t_1)]}\right|_{t_1\to
-\infty}^{t_1=t}
-\left.\left.q^i(A_{12}^{ret}\cdot
A_{21}^{adv})\right|_{[t_1^{ret}(t_2),t_2]}\right|_{t_2\to -\infty}^{t_2=t}.
\end{equation}
Symbolically we denote
\begin{equation}\label{LW}
A_{ba}^{ret}=e_b\frac{v_b[t_b^{ret}(t_a)]}{r_b[t_a,t_b^{ret}(t_a)]},\qquad
A_{ba}^{adv}=e_b\frac{v_b[t_b^{adv}(t_a)]}{r_b[t_a,t_b^{adv}(t_a)]}
\end{equation}
the Li\'enard-Wiechert vector potential of $b$-th charge at point at which
$a$-th one is placed.

\subsection{Radiative part of mixed stress-energy tensor}
In this subsection we find the terms which produce the radiative part
of mixed momentum, either (\ref{p1rad}) or (\ref{p2rad}). We start with
$\varphi$-momentum (\ref{Pirad}) which is then nothing but the mixed
space-time component of the following tensor: \begin{eqnarray} \label{Prad}
{\cal P}^{\mu\nu}_{\rm int, rad}&=&\frac{e_1e_2}{2}\left[
{\hat\Pi }^{\mu\nu}_{12}\left(\frac{\partial^2\sigma}{\partial t_1\partial
t_2}\right)
-\frac{\partial}{\partial t_1}\left(
{\cal D}_1^{\mu}v_2^\nu\frac{\partial^2\sigma}{\partial t_1\partial t_2}
\right)
-\frac{\partial}{\partial t_2}\left(
{\cal D}_2^{\nu}v_1^\mu\frac{\partial^2\sigma}{\partial t_1\partial t_2}
\right)
\right.\nonumber\\
&+&\left.{\hat\Pi}_{21}^{\mu\nu}\left(\frac{\partial^2\sigma}{\partial
t_1\partial t_2}\right)
-\frac{\partial}{\partial t_2}\left(
{\cal D}_2^\mu v_1^\nu\frac{\partial^2\sigma}{\partial t_1\partial t_2}
\right)
-\frac{\partial}{\partial t_1}\left(
{\cal D}_1^\nu v_2^\mu\frac{\partial^2\sigma}{\partial t_1\partial t_2}
\right)
\right].
\end{eqnarray}
To restore $\varphi$-dependent terms leading to this expression, we
insert
\begin{equation}
{\hat\Pi}_{ab}^{\mu\nu}\left(f\right) =
{\hat{\cal T}}_{ab}^{\mu\nu}\left(f\right) -
f\pi_{ab}^{\mu\nu}
\end{equation}
and substitute the right hand side of the third line of
eqs.(\ref{cli}) for $\pi_{ab}^{\mu\nu}$. (Operator ${\hat{\cal
T}}_{ab}^{\mu\nu}$ is defined by eq.(\ref{calTa}).) After
cancellation of like terms we obtain a linear combination of
coefficients (\ref{ABC}). Further we omit the integration over polar
angle and multiply the result on inverse Jacobian $J^{-1}={\rm
q}/(r_1r_2)$. Finally, we obtain the tensor,
\begin{eqnarray} \label{t-rad}
4\pi t^{\mu\nu}\!\!&=&\!\!e_1e_2\left\{
\left(K_1^\mu K_2^\nu +K_2^\mu K_1^\nu\right)\left[
\frac{c_1c_2}{r_1^3r_2^3}\frac{\partial^2\sigma}{\partial t_1\partial t_2}
+\frac{c_1}{r_1^3r_2^2}\frac{\partial^3\sigma}{\partial t_1\partial t_2^2}
+\frac{c_2}{r_1^2r_2^3}\frac{\partial^3\sigma}{\partial t_1^2\partial t_2}
+\frac{1}{r_1^2r_2^2}\frac{\partial^4\sigma}{\partial t_1^2\partial t_2^2}
\right]\right.\nonumber\\
&-&\left.\left(v_1^\mu K_2^\nu +K_2^\mu v_1^\nu\right)\left[
\frac{c_2}{r_1^2r_2^3}\frac{\partial^2\sigma}{\partial t_1\partial t_2}
+\frac{1}{r_1^2r_2^2}\frac{\partial^3\sigma}{\partial t_1\partial t_2^2}
\right]\right.\nonumber\\
&-&\left.\left(K_1^\mu v_2^\nu +v_2^\mu K_1^\nu\right)\left[
\frac{c_1}{r_1^3r_2^2}\frac{\partial^2\sigma}{\partial t_1\partial t_2}
+\frac{1}{r_1^2r_2^2}\frac{\partial^3\sigma}{\partial t_1^2\partial t_2}
\right]\right.\nonumber\\
&+&\left.\left(v_1^\mu v_2^\nu +v_2^\mu v_1^\nu\right)
\frac{1}{r_1^2r_2^2}\frac{\partial^2\sigma}{\partial t_1\partial t_2}
\right\},
\end{eqnarray}
which is symmetric in its indices.

Similarly to the mixed stress-energy tensor (\ref{T12}) itself, its
radiative part,
\begin{equation}\label{Tir}
T_{\rm int,
rad}^{\mu\nu}=t^{\mu\nu}-\frac12\eta^{\mu\nu}t^\alpha{}_\alpha,
\end{equation}
contains the convolution of tensor (\ref{t-rad}),
\begin{eqnarray} \label{t-cnv}
4\pi t^\alpha{}_\alpha\!\!&=&\!\!2e_1e_2\left\{
\frac{c_1c_2}{r_1^3r_2^3}\sigma\frac{\partial^2\sigma}{\partial t_1\partial
t_2}
+\frac{c_1}{r_1^3r_2^2}\frac{\partial}{\partial t_2}
\left(\sigma\frac{\partial^2\sigma}{\partial t_1\partial t_2}\right)
+\frac{c_2}{r_1^2r_2^3}\frac{\partial}{\partial t_1}
\left(\sigma\frac{\partial^2\sigma}{\partial t_1\partial
t_2}\right)\right.\nonumber\\
&+&\left.\frac{1}{r_1^2r_2^2}\frac{\partial^2}{\partial t_1\partial t_2}
\left(\sigma\frac{\partial^2\sigma}{\partial t_1\partial t_2}\right)
+\frac{c_2}{r_1r_2^3}\frac{\partial^2\sigma}{\partial t_1\partial t_2}
+\frac{1}{r_1r_2^2}\frac{\partial^3\sigma}{\partial t_1\partial t_2^2}
\right.\nonumber\\
&+&\left.\frac{c_1}{r_1^3r_2}\frac{\partial^2\sigma}{\partial t_1\partial
t_2}
+\frac{1}{r_1^2r_2}\frac{\partial^3\sigma}{\partial t_1^2\partial t_2}
\right\}.
\end{eqnarray}
Integration over $\varphi$ results in the combination of partial
derivatives in time variables,
\begin{equation}\label{conv}
\int_0^{2\pi}{\rm d}\varphi Jt^\alpha{}_\alpha=e_1e_2\left[
{\hat\Pi}^0\left(\sigma\frac{\partial^2\sigma}{\partial t_1\partial
t_2}\right)
+\frac{\partial}{\partial t_1}
\left(\frac{1}{{\rm q}\|{\sf r}_1\|}\frac{\partial^2\sigma}{\partial
t_1\partial t_2}\right)
+\frac{\partial}{\partial t_2}
\left(\frac{1}{{\rm q}\|{\sf r}_2\|}\frac{\partial^2\sigma}{\partial
t_1\partial t_2}\right)
\right],
\end{equation}
where expressions
\begin{equation}
\|{\sf r}_a\|=\sqrt{[k_a^0-({\mathbf v}_a{\mathbf n}_q)k_a^3]^2-
h^2[{\mathbf v}_a{\mathbf n}_q]^2},
\end{equation}
$a=1,2$, are introduced in \ref{varphi}.

Finally, after a tedious calculations\footnote{Differentiation of
$T_{\rm int, rad}$ is straightforward: one can apply the rule
(\ref{dty}) and combine like terms scaled as $r_1^{-m}r_2^{-n}$;
exponents $m$ and $n$ run from $1$ to $4$ and their sum $3\le m+n\le 7$.}
we derive the identity $\partial_\nu T_{\rm int, rad}^{\mu\nu}=0$.
It means that the radiative part (\ref{Tir}) of mixed energy-momentum tensor
density (\ref{T12})  is conserved off particles' world lines.

It is worth noting that the terms in the first line of expression
(\ref{t-rad}) only belong to the mixed part (\ref{T12}) of the
electromagnetic field stress-energy tensor. The others provide vanishing of
divergence of ${\hat T}_{\rm int, rad}$ as well as reasonable expression
for radiated energy-momentum which escapes to infinity. By means of the
relations
\begin{eqnarray}
v_a^\mu&=&\gamma_a^{-1}u_a^\mu,\qquad
{\dot v}_a^\mu=\gamma_a^{-2}\left(a_a^\mu-\gamma_a^{-1}a_a^0u_a^\mu\right)\\
r_a&=&\gamma_a^{-1}{\sf r}_a,\qquad K_a^\mu={\sf r}_ak_a^\mu,\nonumber\\
c_a&=&\gamma_a^{-2}\left[1+{\sf r}_a(k_a\cdot a_a)+
\gamma_a^{-1}a_a^0{\sf r}_a\right],
\end{eqnarray}
the expression (\ref{t-rad}) can be easily rewritten in manifestly
covariant notations.

\subsection{Radiative part of mixed energy}
Since $\eta^{00}=-1$, the convolution (\ref{conv}) contributes into
zeroth component of radiative part of mixed energy-momentum,
\begin{eqnarray} \label{P0rad}
{\cal P}^0_{\rm int, rad}&=&e_1e_2\left\{
{\hat\Pi}^0\left[\left(k_1^0k_2^0+\frac12\sigma\right)\frac{\partial^2\sigma}{\partial
t_1\partial t_2}\right]\right.\\
&+&\left.\frac12\frac{\partial}{\partial t_1}
\left(\frac{1}{{\rm q}\|{\sf r}_1\|}\frac{\partial^2\sigma}{\partial
t_1\partial t_2}\right)
+\frac12\frac{\partial}{\partial t_2}
\left(\frac{1}{{\rm q}\|{\sf r}_2\|}\frac{\partial^2\sigma}{\partial
t_1\partial t_2}\right)
\right\}.\nonumber
\end{eqnarray}
As could be expected the argument of mixed double derivative,
\begin{equation}\label{dd0}
G^0=\left(k_1^0k_2^0+\frac12\sigma\right)\frac{\partial^2\sigma}{\partial
t_1\partial t_2}{\cal D}^0,
\end{equation}
is not smooth in neighborhoods of end points $\zeta_a\cap\Sigma_t$,
\begin{equation}\label{Delta0}
\Delta{\cal I}^0=-\left.\left.q^0(A_{21}^{ret}\cdot
A_{12}^{adv})\right|_{[t_1,t_2^{ret}(t_1)]}\right|_{t_1\to -\infty}^{t_1=t}
-\left.\left.q^0(A_{12}^{ret}\cdot
A_{21}^{adv})\right|_{[t_1^{ret}(t_2),t_2]}\right|_{t_2\to -\infty}^{t_2=t}.
\end{equation}
Hence the interference rate of radiated energy heavily depends on order of
differentiation, either $\partial^2 G^0/\partial t_1\partial t_2$ or
$\partial^2 G^0/\partial t_2\partial t_1$. Having used the rule
(\ref{d1G0}) we arrive at the expression of type (\ref{p1rad}). Choosing
the rule (\ref{d2G0}) we obtain the expression of type (\ref{p2rad}). The
results can be generalized as follows:
\begin{equation}\label{p0rad}
p_{{\rm rad},ba}^\alpha=
\int_{-\infty}^t{\rm d}t_a\gamma_a^{-1}G_{ba}^\alpha[t_a,t_b^{ret}(t_a)]
+
\int_{-\infty}^{t_a^{ret}(t)}{\rm
d}t_a\gamma_a^{-1}G_{ba}^\alpha[t_a,t_b^{adv}(t_a)],
\end{equation}
where
\begin{eqnarray}\label{fba}
\gamma_a^{-1}G_{ba}^\alpha&=&e_ae_b\left\{\frac{q_{ab}^\alpha(v_a\cdot
v_b)}{r_b^3}\gamma_b^{-2}
+\frac{q_{ab}^\alpha(v_a\cdot v_b)}{r_b^3}(q_{ab}\cdot{\dot v}_b)\right.\\
&+&\left.\frac{q_{ab}^\alpha(v_a\cdot {\dot v}_b)}{r_b^2}-
\frac{v_b^\alpha(v_a\cdot v_b)}{r_b^2}\right\}.\nonumber
\end{eqnarray}

It is convenient to rewrite eq.(\ref{p0rad}) in a manifestly
covariant fashion,
\begin{equation}\label{pabr}
p_{{\rm rad},ab}^\alpha=\int_{-\infty}^{\tau_b}{\rm
d}s_bG_{ab}^\alpha[s_a^{ret}(s_b),s_b]+
\int_{-\infty}^{\tau_b^{ret}(\tau_a)}{\rm d}s_bG_{ab}^\alpha
[s_a^{adv}(s_b),s_b].
\end{equation}
The particles' world lines $\zeta_1$ and $\zeta_2$ are parametrized
by individual proper times $s_1$ and $s_2$, respectively. Upper
limits $\tau_1$ and $\tau_2$ label the points at which $\zeta_1$ and
$\zeta_2$ puncture the observation hyperplane $\Sigma_t$. Two-point
function,
\begin{equation}\label{gab}
G_{ab}^\alpha=e_ae_b\left\{ \frac{q_{ba}^\alpha(u_b\cdot
u_a)}{r_a^3}\left[1+(q_{ba}\cdot
a_a)\right]+\frac{q_{ba}^\alpha(u_b\cdot a_a)}{r_a^2}
-\frac{u_a^\alpha(u_b\cdot u_a)}{r_a^2} \right\},
\end{equation}
is evaluated at points on the world lines of particle $a$ and particle $b$
which are linked by a null ray. $u_a(s_a)$ denotes the normalized
4-velocity of $a$-th particle; 4-acceleration $a_a^\alpha={\rm
d}u_a^\alpha/{\rm d}s_a$.

In contrast with ``one-particle'' contributions to radiated
energy-momentum, the mixed one is not uniquely defined. Having
generalized the expressions (\ref{Deltai}) and (\ref{Delta0}), we
obtain
\begin{equation}\label{Delta}
p_{\rm rad,21}^\alpha-p_{\rm rad,12}^\alpha=
\left.
Q_{21}^\alpha\right|_{s_2\to-\infty}^{s_2=\tau_2}-
\left.
Q_{12}^\alpha\right|_{s_1\to-\infty}^{s_1=\tau_1}\,,
\end{equation}
where
\begin{equation}\label{Qab}
Q^\alpha_{ab}=\left.q^\alpha_{ab}\left(A_{ab}^{adv}\cdot
A_{ba}^{ret}\right)\right|_{[s_a,s_b^{ret}(s_a)]}.
\end{equation}
This $q$-directed null vector is proportional to the scalar product of the
Li\'enard-Wiechert potentials (\ref{LW}).

The difference (\ref{Delta}) depends on the state of particles' motion at
the end points of path integrals. Let us evaluate the net energy-momentum
which escapes to infinity. The integrals over entire world lines should be
substituted for the integrals over the past motion. We suppose that the
particles are asymptotically free. Since Li\'enard-Wiechert potentials fall
off at large distances inversely as the first power of the separation
vector between the charges, the right-hand side of expression (\ref{Delta})
vanishes. Hence the full amount of radiative energy-momentum emitted by
interacting particles does not depend on the method of integration.

\subsection{Radiative part of mixed angular momentum}
When an one-particle problem is considered, bound components of the
stress-energy tensor \cite{Teit} and its torque \cite{LV} contribute
into individual particle's 4-momentum and angular momentum,
respectively. The corrections arise from the regularisation
procedure which involves the Taylor expansion of surface integrals
in which the first two terms lead to the diverging Coulomb
self-energy and the Abraham radiation reaction 4-vector,
respectively. The finite terms depend on the form of the hole that
is cut out from the integration hypersurface to ensure
regularization. The best suited hole must be coordinate-free one.
Teitelboim \cite{Teit} integrate over hyperplane $\sigma(\tau)=\{
y\in {\mathbb M}_{\,4}: u_\mu(\tau)(y^\mu-z^\mu(\tau))=0\}$ which is
orthogonal to the 4-velocity of the charge at point $z(\tau)$ at
which particle's world line punctures $\sigma(\tau)$. It is evident
that the tilted hyperplane together with the future light cone
cutting out the coordinate-free hole (see Refs. \cite[Fig.1]{Teit},
\cite[Fig.1]{LV}, \cite[Fig.5-2]{Rohr}). The considerations lead to
manifestly covariant and structure-independent finite terms of clear
physical sense.

The difficulties associated with the computation of the mixed contribution
(\ref{T12}) are twofold --- to perform the meaningful decomposition of
${\hat T}_{\rm int}$ into bound and radiative parts and to choose an
appropriate surface of integration. The tilted hyperplane which plays
privileged role in the one-particle radiation reaction problem is not
suitable whenever two-body one is considered. Indeed, there is no a
hyperplane which is orthogonal to the world lines of both the particles at
all events. Kosyakov \cite{Kos} constructs a piecewise hypersurface where
a small fragment of a spacelike hyperplane $\Sigma$ is replaced by a
fragment of an orthogonal hyperplane $\sigma_a(\tau_a)=\{y\in{\mathbb
M}_{\,4}:u_{a,\mu}(\tau_a)(y^\mu-z_a^\mu(\tau_a))=0\}$ in the vicinity of
every intersection point. The deformed hyperplane is called {\it locally
adjusted}. But the problem arises how to sew these fragments with $\Sigma$.

Expressions (\ref{pibnd21}), (\ref{pibnd12}), (\ref{p0bnd21}), and
(\ref{p0bnd12}) show that the surface integrals of bound component
of the mixed part of stress-energy tensor depend on the state of
particles' motion in vicinities of intersection points
$\zeta_a\cap\Sigma_t$. Unfortunately, the Taylor expansions of the
divergent components of bound energy-momentum do not lead to
reasonable finite covariant terms. It is because the integration
surface $\Sigma_t$ is tightly connected with the laboratory inertial
frame. This choice yields the coordinate-dependent hole around
$a$-th particle in the point of intersection $\zeta_a\cap\Sigma_t$.
For this reason we assume that an intrinsic structure of a charged
particle is beyond the limits of classical theory \cite{Rohr}. We do
not require any assumptions about the particle structure, its charge
distribution and its size (except that its ``radius'' does not
vanish, although it is too small to be observed). To reconcile the
theory with observation, we assume that a dressed charged particle
possesses {\it finite} 4-momentum and angular momentum.

We face a problem of how the 4-momenta of interacting dressed
charges depend on their individual characteristics such as their
masses, charges, 4-velocities, etc. Valuable information can be
extracted from the radiative part of electromagnetic field's angular
momentum. Indeed, conserved quantities place stringent requirements
on the dynamics of our closed system. They demand that that the
change in radiated momentum and angular momentum should be balanced
by a corresponding change in the individual momenta and angular
momenta of dressed particles. It is felt that analysis of balance
equations yields reasonable expressions. Indeed, it is shown
\cite{Yar03} that the Lorentz-Dirac equation can be derived from the
energy-momentum and angular momentum balance equations. In
\cite{Yar04} the analog of the Lorentz-Dirac equation in six
dimensions is obtained via analysis of 21 conserved quantities which
correspond to the Poincar\'e symmetry of an isolated point particle
coupled with electromagnetic field.

Recall from Section \ref{angul} that the integral of mixed angular
momentum tensor density over $\varphi$ is the combination of partial
derivatives in time variables. To reveal meaningful radiative terms we
integrate eq.(\ref{MM}) over $t_1$ and $t_2$ and apply Teitelboim's
criteria.

The calculation is virtually identical to that presented above, and
we shall not bother with details. The result depends on the method
of integration of mixed double derivatives,
\begin{eqnarray}\label{Mabr}
{\hat M}_{{\rm rad},ab}&=&\int_{-\infty}^{\tau_b}{\rm
d}s_b\left(z_b\wedge G_{ab}^{ret}-e_bu_b\wedge
A_{ab}^{ret}\right)\\
&+&\int_{-\infty}^{\tau_b^{ret}(\tau_a)}{\rm
d}s_b\left(z_b\wedge G_{ab}^{adv}-e_bu_b\wedge
A_{ab}^{adv}\right).\nonumber
\end{eqnarray}
(Symbol $\wedge$ denotes the wedge product.) If we choose the rule
(\ref{d2G0}) we obtain at ${\hat M}_{{\rm rad},21}$. If one prefer
the rule (\ref{d1G0}) they derive ${\hat M}_{{\rm rad},12}$. Direct
calculations show that the difference between these expressions
depends on the momentary state of particles' motion,
\begin{equation}\label{DeltaM}
{\hat M}_{{\rm rad},21}-{\hat M}_{{\rm rad},12}=
\left.z_2\wedge Q_{21}\right|_{s_2\to-\infty}^{s_2=\tau_2}-
\left.z_1\wedge Q_{12}\right|_{s_1\to-\infty}^{s_1=\tau_1}.
\end{equation}

Expression (\ref{Mabr}) for radiated angular momentum satisfies
Teitelboim criteria. Indeed, the integrand is finite and covariant;
the radiation is accumulated with time. Non-covariant bound terms
which are presented in Appendix B are quite different: they depend
on the state of particles' motion at points $\zeta_a\cap\Sigma_t$
and they contain divergences. We assume that the expression
(\ref{Mabr}) is involved in the angular momentum balance equation
explicitly.

\subsection{Radiation of non-accelerated charges}

Let us consider a specific case of very massive particles, such that
$e_1/m_1<<1$ and $e_2/m_2<<1$. Electromagnetic field is too small to
accelerate the charges so that the particles move with constant
velocities. We place the coordinate origin of the Lorentz inertial
frame at point at which the second charge is placed. If ${\mathbf
v}_2=0$, the zeroth component $\gamma_1^{-1}G^0_{21}$ of two-point
function (\ref{fba}) is identically equal to zero. Hence the
radiative energy,
\begin{equation}\label{p0na}
p_{{\rm rad},21}^0= \int_{-\infty}^t{\rm
d}t_1\gamma_1^{-1}G_{21}^0[t_1,t_2^{ret}(t_1)] +
\int_{-\infty}^{t_1^{ret}(t)}{\rm
d}t_1\gamma_1^{-1}G_{21}^0[t_1,t_2^{adv}(t_1)],
\end{equation}
vanishes.

If ${\mathbf v}_2=0$, the space components of this function get
simplified,
\begin{equation}\label{g21na}
\gamma_1^{-1}G_{21}^i=e_1e_2\frac{q^i}{{\rm q}^3}.
\end{equation}
The function evaluated at advanced instant $t_2^{adv}(t_1)$ is just
the function referred to the retarded instant $t_2^{ret}(t_1)$ taken
with opposite sign:
$G_{21}^i[t_1,t_2^{adv}(t_1)]=-G_{21}^i[t_1,t_2^{ret}(t_1)]$.
Because of shift in limits of the retarded and the advanced
integrals the mixed part of radiative $3$-momentum is as follows:
\begin{eqnarray}\label{pina}
p_{{\rm rad},21}^i&=&
\int_{-\infty}^t{\rm d}t_1\gamma_1^{-1}G_{21}^i[t_1,t_2^{ret}(t_1)]
+ \int_{-\infty}^{t_1^{ret}(t)}{\rm
d}t_1\gamma_1^{-1}G_{21}^i[t_1,t_2^{adv}(t_1)]\\
&=&e_1e_2\int_{t_1^{ret}(t)}^t{\rm d}t_1\frac{q^i}{{\rm
q}^3}.\nonumber
\end{eqnarray}
This is vanishingly small quantity. It should be rejected if for no other
reason than that the radiated 4-momentum can not be spacelike 4-vector.

In a like manner we calculate the radiative angular momentum,
\begin{eqnarray}\label{Mina}
{\hat M}_{{\rm rad},21}^{0i}&=&
e_1e_2\int_{t_1^{ret}(t)}^t{\rm
d}t_1\left(t_1\frac{q^i}{{\rm q}^3}+\frac{v_1^i}{\rm q}\right),\\
{\hat M}_{{\rm rad},21}^{ij}&=&0.\nonumber
\end{eqnarray}

If the upper limit of integration $t\to+\infty$, $p_{{\rm
rad},21}^i=0$. Indeed, in this case for each point on $\zeta_1$
there are points on $\zeta_2$ labeled by the retarded and the
advanced instants. Corresponding contributions cancel each other.

Alternative expression for mixed contribution to radiated
energy-momentum,
\begin{equation}\label{pna0}
p_{{\rm rad},12}^\alpha= \int_{-\infty}^{+\infty}{\rm
d}t_2\gamma_2^{-1}G_{12}^\alpha[t_1^{ret}(t_2),t_2] +
\int_{-\infty}^{+\infty}{\rm
d}t_2\gamma_2^{-1}G_{12}^\alpha[t_1^{adv}(t_2),t_2],
\end{equation}
differs from $p_{{\rm rad},21}^\alpha$ on the sum of null vectors,
\begin{equation}
\left.e_1e_2\frac{n_q^\alpha}{{\rm q}\left[1-({\mathbf n}_q{\mathbf
v}_1)\right]}\right|_{t_1\to-\infty}^{t_1\to+\infty}
+\left.e_1e_2\frac{n_q^\alpha}{{\rm q}\left[1+({\mathbf n}_q{\mathbf
v}_1)\right]}\right|_{t_2\to-\infty}^{t_2\to+\infty}\,,
\end{equation}
estimated at limits of integrals. If the particles move with
different velocities, they are asymptotically free. If both the
particles are static, the difference is equal to zero because the
distance ${\rm q}$ between charges does not change.

\section{Equations of motion of radiating charges}\label{eqs}
\setcounter{equation}{0} In this Section we study the
energy-momentum and angular momentum balance equations. We calculate
how much electromagnetic field energy, momentum, and angular
momentum flow across hyperplane $\Sigma_t$. We can do it at a time
$t+\Delta t$. We demand that the change in these quantities be
balanced by a corresponding change in those of the particles, so
that the total energy-momentum,
\begin{equation}\label{P}
P=p_1+p_2+\frac{2e_1^2}{3}\int_{-\infty}^{\tau_1}{\rm
d}s_1a_1^2u_1(s_1) +\frac{2e_2^2}{3}\int_{-\infty}^{\tau_2}{\rm
d}s_2a_2^2u_2(s_2) +p_{\rm int,rad}\,,
\end{equation}
and total angular momentum,
\begin{eqnarray}\label{M}
\hat M&=&z_1\wedge p_1+z_2\wedge p_2\\
&+& \frac{2e_1^2}{3}\int_{-\infty}^{\tau_1}{\rm
d}s_1\left(a_1^2z_1\wedge u_1+u_1\wedge a_1\right)
+\frac{2e_2^2}{3}\int_{-\infty}^{\tau_2}{\rm d}s_2
\left(a_2^2z_2\wedge u_2+u_2\wedge a_2\right) +\hat M_{\rm
int,rad}\,,\nonumber
\end{eqnarray}
are properly conserved.

We suppose that particles' individual 4-momenta $p_1$ and $p_2$ are
already renormalized. The words ``already renormalized'' mean that
momenta contain contributions due to bound component of the stress-energy
tensor density, including its mixed part. While the radiation which
detaches itself from charges and leads an independent existence is
involved explicitly.

In contrast with ``one-particle'' contributions to conserved quantities
(\ref{P}) and (\ref{M}), the mixed ones, (\ref{pabr}) and (\ref{Mabr}),
are not uniquely defined. Our theory faces in the radiation problem a
significant issue: non-uniqueness in determination of radiation sourced by
the mixed part of Maxwell energy-momentum tensor density. To solve the
problem we mix the obtained expressions
\begin{eqnarray}\label{pmix}
p_{\rm int,rad}&=&\kappa p_{\rm rad,21}+(1-\kappa)p_{\rm
rad,12}\,,\\
{\hat M}_{\rm int,rad}&=&\kappa {\hat M}_{\rm
rad,21}+(1-\kappa){\hat M}_{\rm rad,12}\,, \label{Mmix}
\end{eqnarray}
and try to find out the value of constant $\kappa$ that accords with
experience.

Expressions $p_{{\rm rad},ab}$ and ${\hat M}_{{\rm rad},ab}$ are
based on two-point function (\ref{gab}) referred to the points on
the world lines of particle $a$ and particle $b$ which are linked by
a null ray,
\begin{eqnarray}
\sigma&=&-\frac12\left(z_1-z_2\right)^2\\
&=&0.\nonumber
\end{eqnarray}
This implies that a displacement of $z_1(s_1)$ typically induces a
simultaneous displacement of $z_2(s_2)$ because new points
$z_1(s_1+\delta s_1)$ and $z_2(s_2+\delta s_2)$ must also be linked
by a null geodesic,
\begin{equation}
(q_{21}\cdot u_1){\rm d}s_1 + (q_{12}\cdot u_2){\rm d}s_2=0.
\end{equation}
This immediately gives
\begin{equation}\label{dtau}
\frac{{\rm d}s_a}{{\rm d}s_b}=-\frac{(q_{ba}\cdot u_b)}{r_a},
\end{equation}
where $r_a=-(q_{ba}\cdot u_a)$.

By virtue of this expression we compare the retarded and the
advanced integrals involved in eqs.~(\ref{pabr}) and (\ref{Mabr}).
With understanding that functions $s_a^{adv}(s_b)$ and
$s_b^{ret}(s_a)$ are inverses, after some algebra we obtain
\begin{eqnarray}\label{gra}
\!\!\!\!\int_{-\infty}^{\tau_b^{ret}(\tau_a)}{\rm d}s_bG_{ab}^\alpha
[s_a^{adv}(s_b),s_b]&=&\!\!\!\!\left.\int_{-\infty}^{\tau_a}{\rm
d}s_aG_{ba}^\alpha [s_a,s_b^{ret}(s_a)] +
Q^\alpha_{ab}\right|_{s_a\to-\infty}^{s_a=\tau_a},\\
\int_{-\infty}^{\tau_b^{ret}(\tau_a)}{\rm
d}s_b\left(z_b\wedge G_{ab}^{adv}-e_bu_b\wedge
A_{ab}^{adv}\right)&=&\int_{-\infty}^{\tau_a}{\rm
d}s_a\left(z_a\wedge G_{ba}^{ret}-e_au_a\wedge
A_{ba}^{ret}\right)\nonumber\\
&+&\left.\phantom{\int}
z_a\wedge Q_{ab}\right|_{s_a\to-\infty}^{s_a=\tau_a}.
\end{eqnarray}
The relations accord with the right-hand sides of eqs.(\ref{Delta}) and
(\ref{DeltaM}).

Equipped with these relations we rewrite the mixed parts of radiated
energy-momentum (\ref{pabr}) and angular momentum (\ref{Mabr}) as follows:
\begin{eqnarray}\label{pbar}
p_{\rm rad,ab}&=&\left.\int_{-\infty}^{\tau_b}{\rm
d}s_bG_{ab}[s_a^{ret}(s_b),s_b]+
\int_{-\infty}^{\tau_a}{\rm
d}s_aG_{ba}[s_a,s_b^{ret}(s_a)]+Q_{ab}\right|_{s_a\to-\infty}^{s_a=\tau_a},\\
{\hat M}_{\rm rad,ab}&=&\int_{-\infty}^{\tau_b}{\rm
d}s_b\left(z_b\wedge G_{ab}^{ret}-e_bu_b\wedge
A_{ab}^{ret}\right)\\
&+&\left.\int_{-\infty}^{\tau_a}{\rm
d}s_a\left(z_a\wedge G_{ba}^{ret}-e_au_a\wedge
A_{ba}^{ret}\right)+
z_a\wedge Q_{ab}\right|_{s_a\to-\infty}^{s_a=\tau_a}.
\nonumber\label{Mbar}
\end{eqnarray}
Since $Q_{ab}\ne Q_{ba}$, the expressions are not symmetric in indices
$a$ and $b$.

Substituting eqs.(\ref{pmix}) and (\ref{Mmix}) into right-hand sides
of eqs. (\ref{P}) and (\ref{M}), respectively, we obtain the total
energy-momentum,
\begin{equation}
P=\sum_{b=1}^2\left\{\left.
p_b+\frac{2e_b^2}{3}\int_{-\infty}^{\tau_b}{\rm d}s_ba^2_bu_b+
\int_{-\infty}^{\tau_b}{\rm d}s_bG_{ab}^{ret}+
\kappa_bQ_{ba}\right|_{s_b\to-\infty}^{s_b=\tau_b} \right\},
\end{equation}
and angular momentum,
\begin{eqnarray}
{\hat M}&=&\sum_{b=1}^2\left\{
z_b\wedge p_b+
\frac{2e_b^2}{3}\int_{-\infty}^{\tau_b}{\rm d}s_b\left(a^2_bz_b\wedge u_b+
u_b\wedge a_b\right)\right.\\
&+&\left.\left.
\int_{-\infty}^{\tau_b}{\rm d}s_b\left(z_b\wedge
G_{ab}^{ret}-e_bu_b\wedge A_{ab}^{ret}\right)+
\kappa_bz_b\wedge
Q_{ba}\right|_{s_b\to-\infty}^{s_b=\tau_b}\right\}.\nonumber
\end{eqnarray}
(In our notations $\kappa_1:=1-\kappa$ and $\kappa_2:=\kappa$.) The
balance equations are differential consequences of these conserved
quantities. Since the action is not propagated instantaneously, the
balance in a vicinity of the first charge as well as in a
neighborhood of the second charge should be achieved separately,
\begin{eqnarray}\label{padot}
&&{\dot p}_a=-\frac{2e_a^2}{3}a_a^2u_a-G_{ba}^{ret}-\kappa_a{\dot
Q}_{ab}\\
&&u_a\wedge\left(p_a+\frac{2e_a^2}{3}a_a-e_aA_{ba}^{ret}+\kappa_a
Q_{ab}\right)=0.\label{pa}
\end{eqnarray}
(The overdot means the derivation with respect to individual proper
time $\tau_a$.) Solution of six linear equations (\ref{pa}) in four
components of $a$-th 4-momentum contains an arbitrary scalar
function, say $m_a$,
\begin{equation}\label{part}
p_a=m_au_a-\frac{2e_a^2}{3}a_a+e_aA_{ba}^{ret}-\kappa_a Q_{ab}.
\end{equation}
Since $(u_a\cdot a_a)=0$, the scalar product of momentum of $a$-th
particle on its 4-acceleration is as follows:
\begin{equation}\label{apa}
(p_a\cdot a_a)=-\frac{2e_a^2}{3}a_a^2+e_a(A_{ba}^{ret}\cdot a_a)
-\kappa_a (Q_{ab}\cdot a_a).
\end{equation}
Similarly, the scalar product of particle's 4-velocity on the first order
time derivative of particle's 4-momentum (\ref{padot}) is given by
\begin{equation}\label{upd}
({\dot p}_a\cdot u_a)=\frac{2e_a^2}{3}a_a^2-(G_{ba}^{ret}\cdot u_a)
-\kappa_a ({\dot Q}_{ab}\cdot u_a).
\end{equation}
Equipped with the expression (\ref{dtau}) one can derive the
significant relation,
\begin{equation}\label{gFA}
G_{ba}^\alpha=-F_{ba}^\alpha-e_a\frac{{\rm d}A_{ba}^\alpha}{{\rm
d}\tau_a}\,,
\end{equation}
where $F_{ba}^\alpha$ is the well-known Lorentz force of $b$-th charge
acted on $a$-th one. Substituting this into eq.(\ref{upd}) and summing up
(\ref{apa}) and modified (\ref{upd}), we obtain
\begin{equation}\label{pudot}
(p_a\cdot u_a)^\cdot=e_a(A_{ba}^{ret}\cdot u_a)^\cdot
-\kappa_a (Q_{ab}\cdot u_a)^\cdot
\end{equation}
where dot means the $a$-th proper time derivative. On the other hand the
scalar product of 4-momentum (\ref{part}) on $a$-th 4-velocity is written
as
\begin{equation}\label{pua}
(p_a\cdot u_a)=-m_a+e_a(A_{ba}^{ret}\cdot u_a)-\kappa_a (Q_{ab}\cdot
u_a).
\end{equation}
We see clearly that $m_a$ is of constant value. It can be interpreted as
already renormalized mass of $a$-th charged particle.

Finally, we differentiate the expression (\ref{part}) and substitute
it for the left-hand side of eq.(\ref{padot}). After cancellation of
like terms and taking into account eq.(\ref{gFA}) we arrive at the
Lorentz-Dirac equation,
\begin{equation}\label{N2}
m_aa_a=\frac{2e_a^2}{3}\left({\dot a}_a -a_a^2u_a\right) +F_{ba}^{\rm ret}.
\end{equation}
We see that particles' equations of motion do not depend on mixing
parameter $\kappa$.

In terms of kinematical variables conserved quantities of our
particles plus field system looks as follows:
\begin{eqnarray}\label{PP}
P&=&\sum\limits_{a=1}^2\left(m_au_a-\frac{2e_a^2}{3}a_a
+\frac{2e_a^2}{3}\int_{-\infty}^{\tau_a}{\rm
d}s_aa_a^2u_a\right)
-\sum_{a\ne b}\int_{-\infty}^{\tau_a}{\rm d}s_aF_{ba}^{\rm ret},\\
{\hat M}&=&\sum\limits_{a=1}^2\left[z_a\wedge\left(
m_au_a-\frac{2e_a^2}{3}a_a
\right)+\frac{2e_a^2}{3}\int_{-\infty}^{\tau_a}{\rm
d}s_a\left(a_a^2z_a\wedge u_a + u_a\wedge a_a\right)\right]\nonumber\\
&-&\sum_{a\ne b}\int_{-\infty}^{\tau_a}{\rm
d}s_az_a\wedge F_{ba}^{\rm ret}.\label{MMM}
\end{eqnarray}
The work done by Lorentz forces of charges acting on one another exhausts
the radiation reaction due to combination of fields.

Individual 4-momentum (\ref{part}) of $a$-th dressed charged
particle is modified comparing with the well-known Teitelboim's
expression (\ref{p_a}). The bound component ${\hat T}_{\rm int,bnd}$
of mixed part of the electromagnetic field stress-energy tensor
contributes two additional terms: ``immovable core'',
$e_aA_{ba}^{ret}$, of clear physical sense and ``changeable shell'',
$-\kappa_aQ_{ab}$, that heavily depends on mixing parameter
$\kappa$. The corrections are inspired by unavoidable deformation of
bound electromagnetic ``clouds'' due to mutual interaction between
the sources. In my opinion, the changeable term arises due to forced
choice of non-covariant surface of integration (see Subsection 4.4).
For this reason we proclaim the expression
 \begin{equation}\label{ppart}
p_a=m_au_a-\frac{2e_a^2}{3}a_a+e_aA_{ba}^{ret}
\end{equation}
as the only one of true physical meaning.

\section{Discrete symmetries}\label{CPT}
\setcounter{equation}{0} In the past Sections we have emphasized the
importance of the invariance of the action integral (\ref{S}) under
the continuous group of space-time translations and rotations.
According to Noether theorem, these symmetry properties imply
conservation laws, i.~e., those quantities that do not change with
time. In this Section we study symmetry properties of
energy-momentum and angular momentum carried by electromagnetic
field which rely on invariance of (\ref{S}) under discrete
transformation groups.
\subsection{Time inversion}
The transformation of the time inversion ${\cal T}$ is defined by
\cite{Rohr}
\begin{equation}
{\cal T}: y^\mu\mapsto y'^\mu=y_\mu.
\end{equation}
It immediately gives
\begin{equation}
{\cal T}: z_a^\mu(s_a)\mapsto {z'}_a^\mu(s'_a)=z_{a,\mu}(s_a).
\end{equation}
The proper time possesses odd parity \cite{Rohr},
\begin{equation}
{\cal T}: s_a\mapsto s'_a=-s_a.
\end{equation}
The other kinematic quantities then follow easily,
\begin{eqnarray} \label{Tua}
{\cal T}&:& u_a^\mu(s_a)\mapsto {u'}_a^\mu(s'_a)=-u_{a,\mu}(s_a),\\
{\cal T}&:& a_a^\mu(s_a)\mapsto {a'}_a^\mu(s'_a)=a_{a,\mu}(s_a),\nonumber
\end{eqnarray}
etc. The retarded and the advanced instants transform into each
other,
\begin{equation}\label{Ts}
{\cal T}s_a^{ret}(s_b)=s_a^{adv}(s_b),\qquad
{\cal T}s_a^{adv}(s_b)=s_a^{ret}(s_b).
\end{equation}
Inserting eqs.~(\ref{Tua}) and (\ref{Ts}) into two-point function
(\ref{gab}) yields
\begin{eqnarray} \label{Tgab}
{\cal T}G_{ab}^\alpha[s_a^{ret}(s_b),s_b]&=&
-G_{ab,\alpha}[s_a^{adv}(s_b),s_b],\\
{\cal T}G_{ab}^\alpha[s_a^{adv}(s_b),s_b]&=&
-G_{ab,\alpha}[s_a^{ret}(s_b),s_b].\nonumber
\end{eqnarray}

To establish the symmetry properties of $P_{\rm rad}$ and ${\hat
M}_{\rm rad}$ with respect to time inversion, we locate the
observation hyperplane at the distant future,
\begin{eqnarray}\label{Pinf}
P_{\rm rad}&=&
\frac{2e_1^2}{3}\int_{-\infty}^{+\infty}{\rm d}s_1a^2_1u_1+
\frac{2e_2^2}{3}\int_{-\infty}^{+\infty}{\rm d}s_2a^2_2u_2\\
&+&\kappa\int_{-\infty}^{+\infty}{\rm d}s_1\left(
G_{21}^{\rm ret}+G_{21}^{\rm adv}\right)
+(1-\kappa)\int_{-\infty}^{+\infty}{\rm d}s_2\left(
G_{12}^{\rm ret}+G_{12}^{\rm adv}\right),\nonumber\\
{\hat M}_{\rm rad}&=&
\frac{2e_1^2}{3}\int_{-\infty}^{+\infty}{\rm d}s_1\left(a^2_1z_1\wedge u_1+
u_1\wedge a_1\right)
+\frac{2e_2^2}{3}\int_{-\infty}^{+\infty}{\rm d}s_2\left(a^2_2z_2\wedge
u_2+u_2\wedge a_2\right)\nonumber\\
&+&\kappa\int_{-\infty}^{+\infty}{\rm d}s_1\left[
z_1\wedge\left(G_{21}^{\rm ret}+G_{21}^{\rm adv}\right)-
e_1u_1\wedge\left(A_{21}^{\rm ret}+A_{21}^{\rm
adv}\right)\right]\\\label{Minf}
&+&(1-\kappa)\int_{-\infty}^{+\infty}{\rm d}s_2\left[
z_2\wedge\left(G_{12}^{\rm ret}+G_{12}^{\rm adv}\right)-
e_2u_2\wedge\left(A_{12}^{\rm ret}+A_{12}^{\rm adv}\right)\right].\nonumber
\end{eqnarray}
These expressions give the full amount of radiation emitted by interacting
particles.

The transformation of the retarded and the advanced functions
$G_{ab}$ into each other implies that the radiated energy-momentum
and angular momentum are of odd time parity,
\begin{equation}
{\cal T}P^\mu_{\rm rad}=-P_{{\rm rad},\mu}\,,\qquad {\cal T}{\hat
M}^{\mu\nu}_{\rm rad}=-{\hat M}_{{\rm rad},\mu\nu}\,.
\end{equation}

\subsection{Space inversion}
This operation is defined by \cite{Rohr}
\begin{equation}
{\cal P}: y^\mu\mapsto y'^\mu=-y_\mu.
\end{equation}
Correspondingly,
\begin{equation} \label{Pz}
{\cal P}: z_a^\mu(s_a)\mapsto {z'}_a^\mu(s'_a)=-z_{a,\mu}(s_a).
\end{equation}
The proper time $s_a$ remains invariant. These transformation properties
imply that that particles' 4-velocities and 4-accelerations change as
follows:
\begin{eqnarray} \label{Pua}
{\cal P}&:& u_a^\mu(s_a)\mapsto {u'}_a^\mu(s'_a)=-u_{a,\mu}(s_a),\\
{\cal P}&:& a_a^\mu(s_a)\mapsto
{a'}_a^\mu(s'_a)=-a_{a,\mu}(s_a).\nonumber
\end{eqnarray}
Since the retarded and the advanced instants remain invariant, the
basic two-point function (\ref{gab}) transforms analogously,
\begin{equation}
{\cal P}G_{ab}^\alpha=-G_{ab,\alpha}.
\end{equation}
Substituting this into eqs.(\ref{Pinf}) and (\ref{Minf}) and using the
relations (\ref{Pz}) and (\ref{Pua}) yields
\begin{equation}
{\cal P}P^\mu_{\rm rad}=-P_{{\rm rad},\mu}\,,\qquad {\cal P}{\hat
M}^{\mu\nu}_{\rm rad}={\hat M}_{{\rm rad},\mu\nu}\,.
\end{equation}

\subsection{Reciprocity of particles 1 and 2}
When a closed system of two identical charges is considered, radiative
conserved quantities should be symmetric in indices $1$ and $2$ that label
the particles. Having interchanged these indices in function (\ref{gab})
we obtain
\begin{eqnarray}\label{g1i2}
\left.G^\alpha_{21}[s_1,s_2^{ret}(s_1)]\right|_{1\leftrightarrow 2}&=&
G^\alpha_{12}[s_1^{ret}(s_2),s_2],\\
\left.G^\alpha_{21}[s_1,s_2^{adv}(s_1)]\right|_{1\leftrightarrow 2}&=&
G^\alpha_{12}[s_1^{adv}(s_2),s_2].\nonumber
\end{eqnarray}
From the reciprocity relations we see clearly that mixed parameter
$\kappa$ in expressions (\ref{Pinf}) and (\ref{Minf}) should be equal to
$1/2$. Choosing the linear superposition
\begin{eqnarray}
p_{\rm int,rad}&=&\frac12\left(p_{\rm rad, 21}+p_{\rm rad,
12}\right),\\
{\hat M}_{\rm int,rad}&=&\frac12\left({\hat M}_{\rm rad, 21}+ {\hat
M}_{\rm rad, 12}\right),
\end{eqnarray}
we restore invariance of radiated energy-momentum and angular
momentum with respect to reciprocity of particles $1$ and $2$.

\section{Conclusions}\label{cncl}
\setcounter{equation}{0}

The present paper is devoted to study of phenomena of emission and
propagation of energy in classical electrodynamics. The field in action
(\ref{S}) has its own uncountably infinite degrees of freedom. Variation of
(\ref{S}) yields Maxwell's equations with point-like sources and equations
of motions of particles interacting through the medium of the field. The
problem then becomes one of mutual determination: the field is determined
by the charged particles and their motion, and the motion of the charges is
determined by the field.

In this paper we study interference of outgoing electromagnetic
waves in a hyperplane $\Sigma_t=\{y\in {\mathbb M}_{\,4}: y^0=t\}$
associated with an unmoving inertial observer. We calculate how much
electromagnetic field energy, momentum, and angular momentum flow
across this hyperplane. Surface integration of the stress-energy
tensor (\ref{T}) over $\Sigma_t$ reduces field's uncountably
infinite degrees of freedom. After the renormalization procedure we
arrive at the action at a distance theory \cite{WF1,WF2} where
particles interact directly with one another. The fields in
resulting expressions (\ref{PP}) and (\ref{MMM}) do not have degrees
of freedom of their own: they are functionals of particle paths.
Following Ref.~\cite{HN}, we refer to them as direct particle
fields.

Starting with the retarded Li\'enard-Wiechert fields, after integration we
arrive at the retarded and the advanced direct particle fields. The
retarded and the advanced instants arise naturally as the end points of
interference integrals (\ref{int1}) and (\ref{int2}). But the retarded
causality is not violated because the advanced instants as well as the
retarded ones are before the fixed observation moment $t$. Direct particle
fields referred to advanced instants do not describe neither incoming
radiation nor converging electromagnetic waves.

Nevertheless, the retarded and the advanced quantities transform
into each other under the influence of inversion of time axes. To
demonstrate the invariance of electromagnetic field's
energy-momentum and angular momentum with respect to time inversion,
we locate the observation surface $\Sigma_t$ at distant future. We
show that the full amount of radiation emitted by a closed system of
two interacting charged particles is invariant with respect to past
and future as well as with respect to inversion of space axes.

The part of Noether quantities which escapes to infinity is defined
by basic two-point function (\ref{gab}). It is equal to the sum of
Lorentz force $F_{ab}$ of $a$-th charge acted on $b$-th one and the
total time derivative of direct Li\'enard-Wiechert potential
$A_{ab}$, taken with opposite sign (see eq.(\ref{gFA})). In the
specific case of very massive particles, such that $e_1/m_1<<1$ and
$e_2/m_2<<1$, the mixed energy-momentum $p_{\rm int,rad}$ vanishes
as could be expected for nonaccelerated charges. Indeed, let us
consider static charge $b$ at a coordinate origin. Zeroth component
$F_{ba}^0$, either retarded or advanced, is potential one: its
integral over indicated portion of $\zeta_a$ cancels the change of
only nontrivial $A_{ba}^0$. Space components $F_{ba}^i$ of the
retarded and the advanced Lorentz forces compensate each other.
Having performed a trivial Lorentz transformation we extend the
statement on a motion with constant velocity. Radiative angular
momentum possesses analogous properties.

To derive the equations of motion of interacting particles we
compare flows of energy-momentum and angular momentum through very
close hyperplanes $\Sigma_t$ and $\Sigma_{t+\Delta t}$. Having
balanced particles' individual characteristics and corresponding
quantities carried by electromagnetic field, we obtain the
well-known Lorentz-Dirac equation of motion of charged particle in
the retarded field of the other charge where the self-action is
taken into account. Since the Lorentz-Dirac equation possesses
pathological solutions, such as runaway solutions or
preaccelerations, the authors \cite{Sp,Rhr1,Pr} propose the
Landau-Lifshitz equation \cite[\S 76]{LL} as more satisfactory
alternative. Spohn \cite{Sp} showed that a solution of the
Lorentz-Dirac equation which does not satisfy the Landau-Lifshitz
equation is of the runaway type. Moreover, Landau-Lifshitz equation
does not permit runaway solutions or preacceleration \cite{Pr}.

Obtained expressions for radiated energy-momentum and angular
momentum (including interference effects) are valuable also for a
strictly prescribed motion of particles under the influence of a
very powerful external force. It is necessary to compare them with
corresponding results for circling charges \cite{VR,RV}. It would be
interesting to consider the specific case when the charged particles
are a very close to each other. Since the electromagnetic field
satisfies the superposition principle, the models either an extended
object consisting of $N$ point charges or a continuous charge
distribution are based on dynamics of two-body system
\cite{OR1,OR2}.

\section*{Acknowledgments}
I am grateful to  V.I. Tre\-tyak for continuous encouragement
and for a helpful reading of this manuscript. I would like to thank
A. Du\-vi\-ryak for many useful discussions.

\setcounter{subsubsection}{0}
\renewcommand{\thesubsubsection}{Appendix \Alph{subsubsection}}
\subsubsection{Integration over angular variable}\label{varphi}
\renewcommand{\theequation}{\Alph{subsubsection}.\arabic{equation}}
\setcounter{equation}{0}

To calculate the mixed rates of energy-momentum (\ref{p_int}) and
angular momentum (\ref{M_int}) carried by the electromagnetic field,
we should first perform the integration over angle. When facing this
problem it is convenient to mark out $\varphi$-dependent terms in
expressions under the integral sign. In the Maxwell energy-momentum
tensor density we distinguish the second-order differential operator
(\ref{calTa}) with $\varphi$-dependent coefficients (\ref{ABC}). It
can be decomposed into a combination of partial derivatives in time
variables ${\hat \Pi}_a$ given by eq.(\ref{Pap}) and tail $\pi_a$ of
the type in eq.(\ref{pia}).

This Appendix is concerned with the computation of the tails. Equipped
with them we express the aforementioned integrand as a combination of
partial derivatives in $t_1$ and $t_2$.

To implement this strategy we must first integrate the coefficients
(\ref{ABC}) over the angle variable. We start with the simplest one,
\begin{equation}\label{BD}
{\cal D}^a=\frac{1}{2\pi}\int_0^{2\pi}{\rm d}\varphi\frac{a}{{\rm
q}r_1r_2},
\end{equation}
where numerator $a$ is equal to $1$ or $K_1^\mu$. Our task is to rewrite
the integrand as a sum of term with denominator $r_1$ and term with
denominator $r_2$. To do it we introduce a new layer of mathematical
formalism and develop convenient technique.

Let $V$ be the vector space such that ${\mathbf i}_0, {\mathbf i}_1$,
and ${\mathbf i}_2$ is its linear basis. We shall use
$\eta_{\alpha\beta}={\rm diag}(-1,1,1)$ and its inverse
$\eta^{\alpha\beta}={\rm diag}(-1,1,1)$ to lower and raise indices,
respectively. We introduce the pairing
\begin{eqnarray}
(\cdot)&:&V\times V\to \mathbb R\\
&& ({\sf a}\cdot{\sf b})\mapsto\eta_{\alpha\beta}{\sf a}^\alpha{\sf
b}^\beta,\nonumber
\end{eqnarray}
which will be called the ``scalar product''.

We introduce null vector $n^\alpha=(1,\sin\varphi,\cos\varphi)$ which
belongs to the vector space $V$. We express the $\varphi$-dependent
constructions
\begin{equation}\label{rc}
r_a=-(K_a\cdot v_a), \qquad c_a=\gamma_a^{-2}+(K_a\cdot {\dot v}_a)
\end{equation}
as the scalar products $-({\sf r_a}\cdot n)$ and $({\sf c_a}\cdot
n)$, respectively. We shall use {\sf sans-serif} letters for the
components of timelike three-vectors ${\sf r}_a\in V$ and ${\sf
c}_a\in V$,
\begin{eqnarray}\label{r0}
{\sf r}_a^0&=&k_a^0-({\mathbf v}_a{\mathbf n}_{\rm q})k_a^3,\qquad
{\sf r}_a^1=hv_a{}^i\omega_i{}^1, \qquad
{\sf r}_a^2=hv_a{}^i\omega_i{}^2;\\
{\sf c}_a^0&=&-\gamma_a^{-2}-({\mathbf\dot v}_a{\mathbf n}_{\rm
q})k_a^3,\qquad {\sf c}_a^1=h{\dot v}_a{}^i\omega_i{}^1,\qquad {\sf
c}_a^2=h{\dot v}_a{}^i\omega_i{}^2,\label{c0}
\end{eqnarray}
where orthogonal matrix,
\begin{equation}\label{om}
\hat\omega=\left(
\begin{array}{ccc}
\cos\varphi_q&-\sin\varphi_q&0\\
\sin\varphi_q&\cos\varphi_q&0\\
0 & 0 & 1
\end{array}
\right)
\left(
\begin{array}{ccc}
\cos\vartheta_q&0&\sin\vartheta_q\\
0 & 1 & 0\\
-\sin\vartheta_q&0&\cos\vartheta_q
\end{array}
\right),
\end{equation}
is constructed from components of the relative position 3-vector
${\mathbf q}={\mathbf z}_1-{\mathbf z}_2$.
The numerator in eq.(\ref{BD}) is the scalar product
$({\sf a}\cdot n)=-{\sf a}^0+{\sf a}^1\sin\varphi +{\sf a}^2\cos\varphi$;
it is equal to $1$ if vector ${\sf a}=(-1,0,0)$.

We introduce the dual space of one-forms, say $W$, with basis
${\hat\omega}^0,{\hat\omega}^1$ and ${\hat\omega}^2$, such that
${\hat\omega}^\mu({\mathbf i}_\nu)=\delta_\nu^\mu$, where ${\mathbf
i}_0, {\mathbf i}_1, {\mathbf i}_2$ constitute the basis of $V$. The
wedge product ${\hat L}={\hat a}\wedge{\hat b}$ of two one forms
${\hat a}$ and ${\hat b}$ constitutes two-form,
\begin{equation}
{\hat
L}=\left(a_0b_1-a_1b_0\right){\hat\omega}^0\wedge{\hat\omega}^1+
\left(a_0b_2-a_2b_0\right){\hat\omega}^0\wedge{\hat\omega}^2+
\left(a_1b_2-a_2b_1\right){\hat\omega}^1\wedge{\hat\omega}^2.
\end{equation}
We introduce dual three-vector ${\sf L}=^*\!\!{\hat L}$ with components
\begin{eqnarray}\label{hatL}
{\sf L}^\alpha&=&\frac{1}{2!}\varepsilon^{\alpha\beta\gamma}L_{\beta\gamma}
\\
&=&\varepsilon^{\alpha\beta\gamma}a_\beta b_\gamma.\nonumber
\end{eqnarray}
$\varepsilon^{\alpha\beta\gamma}$ denotes the Ricci symbol in three
dimensions,
\begin{equation}\label{eps}
\varepsilon^{\alpha\beta\gamma}=\left\{
\begin{array}{ccccccccc}
1 &{\rm when}& \alpha\beta\gamma& {\rm is}& {\rm an}& {\rm even}&
{\rm permutation}& {\rm of}& 0,1,2\\
-1& {\rm when}& \alpha\beta\gamma& {\rm is}& {\rm an}& {\rm odd}&
{\rm permutation}& {\rm of}& 0,1,2\\
0& {\rm otherwise}.&&&&&&&
\end{array}
\right.
\end{equation}

We raise indices in eq.(\ref{hatL}) and define the vector product of
two vectors, ${\sf a}$ and ${\sf b}$,
\begin{equation}\label{sfL}
{\sf L}^\alpha=\varepsilon^\alpha{}_{\mu\nu}{\sf a}^\mu{\sf b}^\nu.
\end{equation}
Tensor,
\begin{equation}\label{ep}
\varepsilon^\alpha{}_{\mu\nu}=\varepsilon^{\alpha\beta\gamma}\eta_{\beta\mu}
\eta_{\gamma\nu},
\end{equation}
has the components
\begin{equation}\label{epp}
\varepsilon^0{}_{\mu\nu}=\varepsilon^{0\mu\nu},\qquad
\varepsilon^1{}_{\mu\nu}=-\varepsilon^{1\mu\nu},\qquad
\varepsilon^2{}_{\mu\nu}=-\varepsilon^{2\mu\nu}.
\end{equation}

Now we calculate the double vector product,
\begin{equation}
[{\sf A}[{\sf B}{\sf C}]]^\alpha=
\varepsilon^\alpha{}_{\beta\gamma}{\sf A}^\beta\varepsilon^\gamma{}_{\mu\nu}
{\sf B}^\mu{\sf C}^\nu.
\end{equation}
Since
\begin{equation}
\varepsilon^\alpha{}_{\beta\gamma}\varepsilon^\gamma{}_{\mu\nu}=
-\delta^\alpha{}_\mu\eta_{\beta\nu}+\delta^\alpha{}_\nu\eta_{\beta\mu},
\end{equation}
we arrive to the unusual rule,
\begin{equation}\label{vpr}
[{\sf A}[{\sf B}{\sf C}]]=-{\sf B}({\sf A}\cdot{\sf C})+ {\sf
C}({\sf A}\cdot{\sf B}),
\end{equation}
instead of the well-known law acting in space with Euclidean metric.

To simplify the denominator $r_1r_2$ in the integrand of
eq.(\ref{BD}) as much as possible, we rewrite $2\pi$-periodic
functions $r_a=-{\sf r}_{a,0}-{\sf r}_{a,1}\sin\varphi-{\sf
r}_{a,2}\cos\varphi$ as follows:
\begin{equation}\label{Bra}
r_a=-{\sf r}_{a,0}-\rho_a\sin(\varphi+\phi_a),\qquad
\rho_a=\sqrt{{\sf r}_{a,1}^2+{\sf r}_{a,2}^2}.
\end{equation}
(We recall that $r_a$ is the scalar products $({\sf r}_a\cdot n)$ taken
with opposite sign, components ${\sf r}_a^\mu$ are given
by eqs.(\ref{r0}).) Shift in argument of harmonic function is determined
by the relations
\begin{equation}
{\sf r}_{a,1}=\rho_a\cos\phi_a,\qquad
{\sf r}_{a,2}=\rho_a\sin\phi_a.
\end{equation}
After some algebra we rewrite the integrand of eq.(\ref{BD}) as the
following sum:
\begin{equation}\label{BI}
\frac{a}{r_1r_2}=\frac{A_{12}^a+C_{12}^a\rho_1\cos(\varphi+\phi_1)}{r_1}+
\frac{A_{21}^a-C_{12}^a\rho_2\cos(\varphi+\phi_2)}{r_2},
\end{equation}
where $a=({\sf a}\cdot n)$. Coefficients $A_{12}^a, A_{21}^a$ and
$C_{12}^a$ satisfy the vector equation
\begin{equation}\label{be}
- A_{12}^a{\sf r}_2 - A_{21}^a{\sf r}_1 + C_{12}^a{\sf L}_{12}={\sf a},
\end{equation}
where by ${\sf r}_1$ and ${\sf r}_2$ we mean three-vectors with components
in eq.(\ref{r0}) and ${\sf L}_{12}=[{\sf r}_1{\sf r}_2]$.

To solve equation (\ref{be}) we postmultiply it on the vector
product $[{\sf r}_1{\sf L}_{12}]$, then on the vector product $[{\sf
r}_2{\sf L}_{21}]$, and, finally, on ${\sf L}_{12}$. After some
algebra we obtain
\begin{equation}\label{AAC}
A_{12}^a=\frac{\left([{\sf a}{\sf r}_1]\cdot{\sf L}_{12}\right)}{D_{12}},
\qquad
A_{21}^a=\frac{\left([{\sf a}{\sf r}_2]\cdot{\sf L}_{21}\right)}{D_{21}},
\qquad
C_{12}^a=\frac{\left({\sf a}\cdot{\sf L}_{12}\right)}{D_{12}},
\end{equation}
where the denominator $D_{12}=({\sf L}_{12}\cdot{\sf L}_{12})$ is symmetric
in its indices.

Substituting eq.(\ref{BI}) into eq.(\ref{BD}) and using the identities
\begin{eqnarray}\label{B1}
\frac{1}{2\pi}\int_0^{2\pi}
\frac{{\rm d}\varphi}{{\sf r}_a^0-\rho_a\sin(\varphi+\phi_a)}&=&
\frac{1}{\sqrt{({\sf r}_a^0)^2-\rho_a^2}}\\
&=&\frac{1}{\|{\sf r}_a\|},\nonumber\\
\frac{1}{2\pi}\int_0^{2\pi}{\rm d}\varphi
\frac{\cos(\varphi+\phi_a)}{{\sf r}_a^0-\rho_a\sin(\varphi+\phi_a)}&=&0
\nonumber
\end{eqnarray}
yields
\begin{equation}\label{Dfin}
{\cal D}^a=\frac{A_{12}^a}{{\rm q}\|{\sf
r}_1\|}+\frac{A_{21}^a}{{\rm q}\|{\sf r}_2\|}
\end{equation}
after integration over $\varphi$.

Now we turn to the calculation of the coefficient
\begin{equation}\label{B}
{\cal B}^a=\frac{1}{2\pi}\int_0^{2\pi}{\rm d}\varphi\frac{ac_2}{{\rm
q}r_1(r_2)^2}.
\end{equation}
Equipped with the relations in eq.(\ref{BI}) we rewrite the integrand as a
sum of terms which are proportional to the $1/r_1$, $1/r_2$, and
$1/(r_2)^2$, respectively. Using the identities
\begin{eqnarray}\label{B2}
\frac{1}{2\pi}\int_0^{2\pi}
\frac{{\rm d}\varphi}{\left[{\sf
r}_a^0-\rho_a\sin(\varphi+\phi_a)\right]^2}&=&\frac{{\sf r}_a^0}{\|{\sf
r}_a\|^3},\\
\frac{1}{2\pi}\int_0^{2\pi}{\rm d}\varphi
\frac{\cos(\varphi+\phi_a)}{\left[{\sf r}_a^0-\rho_a\sin(\varphi+
\phi_a)\right]^2}&=&0
\nonumber
\end{eqnarray}
and taking into account the relations in eq.(\ref{B1}) gives
\begin{eqnarray}\label{Bfin}
{\cal B}^a&=&-\frac{1}{{\rm q}\|{\sf r}_2\|^3}
\frac{({\sf a}\cdot{\sf r}_2)({\sf c}_2\cdot{\sf r}_2)}{D_{21}}
({\sf r}_2\cdot{\sf r}_1)
+\frac{1}{{\rm q}\|{\sf r}_2\|}\left[A_{12}^{c_2}A_{21}^a+
A_{12}^aA_{21}^{c_2}-\frac{({\sf a}\cdot{\sf c}_2)({\sf r}_1\cdot{\sf r}_2)}
{D_{21}}\right]\nonumber\\
&+&\frac{1}{{\rm q}\|{\sf r}_1\|}\left[2A_{12}^aA_{12}^{c_2}-\frac{([{\sf
a}{\sf r}_1]\cdot[{\sf c}_2{\sf r}_1])}
{D_{12}}\right].
\end{eqnarray}
The resulting expression for the term
\begin{equation}\label{BC}
{\cal C}^a=\frac{1}{2\pi}\int_0^{2\pi}{\rm d}\varphi\frac{ac_1}{{\rm
q}(r_1)^2r_2}
\end{equation}
can be obtained by interchanging indices 1 and 2 in the right-hand side of
eq.(\ref{Bfin}).

After a routine computation based on the repeated usage of relation
(\ref{BI}), we find the most complicate term,
\begin{eqnarray}\label{BA}
{\cal
A}^a&=&\frac{1}{2\pi}\int_0^{2\pi}{\rm d}\varphi\frac{ac_1c_2}{{\rm
q}(r_1)^2(r_2)^2}\\
&=&\frac{B_{12}}{{\rm q}\|{\sf r}_1\|}+\frac{B_{21}}{{\rm q}\|{\sf
r}_2\|}+
J_1\frac{({\sf a}\cdot{\sf r}_1)}{{\rm q}\|{\sf r}_1\|^3}+
J_2\frac{({\sf a}\cdot{\sf r}_2)}{{\rm q}\|{\sf r}_2\|^3},\nonumber
\end{eqnarray}
where
\begin{eqnarray}
J_1&=&2A_{12}^{c_1}A_{12}^{c_2}-
\frac{([{\sf c}_1{\sf r}_1]\cdot[{\sf c}_2{\sf r}_1])}{D_{12}},\\
B_{12}&=&3A_{12}^aA_{12}^{c_1}A_{21}^{c_2}+
3A_{12}^aA_{12}^{c_2}A_{21}^{c_1}+
2A_{12}^{c_1}A_{12}^{c_2}A_{21}^a+
A_{12}^a\left\{
\frac{([{\sf c}_1{\sf r}_1]\cdot[{\sf c}_2{\sf r}_2])}{D_{12}}+
\frac{([{\sf c}_1{\sf r}_2]\cdot[{\sf c}_2{\sf r}_1])}{D_{12}}
\right\}\nonumber\\
&-&A_{21}^{c_1}\frac{([{\sf c}_2{\sf r}_1]\cdot[{\sf a}{\sf r}_1])}{D_{12}}
-A_{21}^{c_2}\frac{([{\sf c}_1{\sf r}_1]\cdot[{\sf a}{\sf r}_1])}{D_{12}}
+A_{12}^{c_1}\frac{([{\sf c}_2{\sf r}_1]\cdot[{\sf a}{\sf r}_2])}{D_{12}}
+A_{12}^{c_2}\frac{([{\sf c}_1{\sf r}_1]\cdot[{\sf a}{\sf r}_2])}{D_{12}},
\nonumber
\end{eqnarray}
and the others, $B_{21}$ and $J_2$, can be obtained via interchanging
indices $1$ and $2$.

We now turn to the differentiation of coefficient (\ref{Dfin}) with
respect to time variables $t_1$ and $t_2$. Having substituted ${\sf
o}=(-1,0,0)$ for ${\sf a}$ in the expressions (\ref{Dfin}),
(\ref{Bfin}), and (\ref{BA}) we obtain the terms ${\cal D}^0$,
${\cal B}^0$ and ${\cal A}^0$, respectively. The remaining term,
${\cal C}^0$, can be obtained from ${\cal B}^0$ via reciprocity. The
calculations are based on the relations obtained via differentiation
of third components $k_a^3$ and $h^2$, i.e. the square of radius of
the circle $C(O,h)=S_1\cap S_2$,
\begin{eqnarray}
\frac{\partial k_a^3}{\partial t_a}&=&-\left[(-1)^a\frac{r_a^0}{{\rm q}}+
({\mathbf n}_{\rm q}{\mathbf v}_a)\right],\qquad
\frac{\partial k_a^3}{\partial t_b}=(-1)^a\frac{r_b^0}{{\rm q}},\nonumber\\
\frac{\partial h^2}{\partial t_1}&=&-2k_2^3\frac{r_1^0}{{\rm q}},\qquad
\frac{\partial h^2}{\partial t_2}=2k_1^3\frac{r_2^0}{{\rm q}}.\label{h2}
\end{eqnarray}
They immediately give
\begin{eqnarray}
\frac{\partial {\sf r}_a^0}{\partial t_a}&=&{\sf c}_a^0
+\frac{(-1)^a}{{\rm q}}\left[({\mathbf v}_a{\mathbf n}_{\rm q}){\sf
r}_a^0+
[{\mathbf v}_a{\mathbf n}_{\rm q}]^2k_b^3\phantom{\frac11}\!\!\!\!\right],
\\
\frac{\partial {\sf r}_a^0}{\partial t_b}&=&
\frac{(-1)^b}{{\rm q}}\left[({\mathbf v}_a{\mathbf n}_{\rm q}){\sf r}_b^0+
([{\mathbf v}_a{\mathbf n}_{\rm q}][{\mathbf v}_b{\mathbf n}_{\rm
q}])k_a^3
\phantom{\frac11}\!\!\!\!\right],
\nonumber
\end{eqnarray}
and eventually give
\begin{eqnarray}\label{dra}
\frac{\partial ({\sf r}_a\cdot{\sf r}_a)}{\partial t_a}&=&
2\left[({\sf r}_a\cdot{\sf c}_a)
+(-1)^a\frac{({\mathbf v}_a{\mathbf n}_{\rm q})}{{\rm q}}({\sf
r}_a\cdot{\sf r}_a)
\right],\\
\frac{\partial ({\sf r}_a\cdot{\sf r}_a)}{\partial t_b}&=&
2(-1)^b\left[\frac{({\mathbf v}_a{\mathbf n}_{\rm q})}{{\rm q}}({\sf
r}_a\cdot{\sf r}_b)+\frac{k_a^3}{{\rm q}}\frac{{\sf r}_b^0({\sf
r}_a\cdot{\sf r}_a)-
{\sf r}_a^0({\sf r}_a\cdot{\sf r}_b)}{h^2}
\right],
\nonumber\\
\frac{\partial ({\sf r}_b\cdot{\sf r}_a)}{\partial t_b}&=&
({\sf r}_a\cdot{\sf c}_b)+
(-1)^b\left[
\frac{({\mathbf v}_b{\mathbf n}_{\rm q})}{{\rm q}}({\sf r}_b\cdot{\sf
r}_a)
+\frac{({\mathbf v}_a{\mathbf n}_{\rm q})}{{\rm q}}({\sf r}_b\cdot{\sf
r}_b)
+\frac{k_a^3}{{\rm q}}\frac{{\sf r}_b^0({\sf r}_b\cdot{\sf r}_a)-
{\sf r}_a^0({\sf r}_b\cdot{\sf r}_b)}{h^2}\right].
\nonumber
\end{eqnarray}
We use Latin indices $a$ and $b$ which run from $1$ to $2$ ($a\ne b$).
We use bold script for conventional three-velocities $v_a^i={\rm
d}z_a^i/{\rm d}t_a$ and unit three-vector $n_{\rm q}^i=q^i/{\rm q}$ in
${\mathbf q}$-direction. By $({\mathbf v}_b{\mathbf n}_{\rm q})$ and
$[{\mathbf v}_b{\mathbf n}_{\rm q}]$ we denote the conventional scalar
product and cross product of these vectors, respectively.

Usage of these relations allows us to calculate the derivatives of
coefficients $A_{ab}^0$:
\begin{eqnarray}
\frac{\partial A_{ab}^0}{\partial t_a}&=&A_{ab}^0A_{ba}^{c_a}+
A_{ab}^{c_a}A_{ba}^0+\frac{([{\sf o}{\sf r}_a]\cdot[{\sf c}_a{\sf
r}_b])}{D_{ab}},\\
\frac{\partial A_{ab}^0}{\partial t_b}&=&
2A_{ab}^{c_b}A_{ab}^0-\frac{([{\sf o}{\sf r}_a]\cdot[{\sf c}_b{\sf
r}_a])}{D_{ab}}+(-1)^b\frac{k_a^3}{{\rm q}h^2}
\left({\sf o}^0+{\sf r}_2^0A_{12}^0+{\sf r}_1^0A_{21}^0
\right)\nonumber\\
&+&(-1)^a\left(A_{ab}^0\frac{({\mathbf n}_{\rm q}{\mathbf v}_b)}{{\rm q}}
+A_{ba}^0\frac{({\mathbf n}_{\rm q}{\mathbf v}_a)}{{\rm
q}}\right).\nonumber
\end{eqnarray}
Substituting these into equality
\begin{equation}
\frac{\partial{\cal D}^0}{\partial t_a}=\frac{\partial}{\partial t_a}
\left(
\frac{A_{12}^0}{{\rm q}\|{\sf r}_1\|}+\frac{A_{21}^0}{{\rm q}\|{\sf
r}_2\|}
\right)
\end{equation}
and using the identities
\begin{eqnarray}\label{d1ra}
\frac{\partial}{\partial t_a}\left(\frac{1}{\|{\sf r}_a\|}
\right)&=&\frac{({\sf c}_a\cdot{\sf r}_a)}{\|{\sf r}_a\|^3}
-(-1)^a\frac{1}{\|{\sf r}_a\|}\frac{({\mathbf n}_{\rm q}{\mathbf
v}_a)}{{\rm q}},
\\
\frac{\partial}{\partial t_b}\left(
\frac{1}{\|{\sf r}_a\|}
\right)&=&(-1)^b\frac{({\sf r}_a\cdot{\sf r}_b)}{\|{\sf r}_a\|^3}
\frac{({\mathbf n}_{\rm q}{\mathbf v}_a)}{{\rm q}}+(-1)^a
\frac{k_a^3}{{\rm q}\|{\sf r}_a\|^3}
\frac{{\sf r}_a^0({\sf r}_a\cdot{\sf r}_b)
-{\sf r}_b^0({\sf r}_a\cdot{\sf r}_a)}{h^2}\nonumber
\end{eqnarray}
yields
\begin{equation}\label{D0}
\frac{\partial{\cal D}^0}{\partial t_1}={\cal C}^0+
\frac{k_2^0({\mathbf n}_{\rm q}{\mathbf v}_2)-
k_2^3{\mathbf v}_2^2}{{\rm q}^2\|{\sf r}_2\|^3},
\qquad
\frac{\partial{\cal D}^0}{\partial t_2}={\cal B}^0
-\frac{k_1^0({\mathbf n}_{\rm q}{\mathbf v}_1)-
k_1^3{\mathbf v}_1^2}{{\rm q}^2\|{\sf r}_1\|^3}.
\end{equation}
Further we calculate the partial derivative $\partial{\cal
C}^0/\partial t_2$, subtract it from ${\cal A}^0$, and prove the
identity
\begin{equation}\label{pi0}
{\cal A}^0-\frac{\partial{\cal C}^0}{\partial t_2}=
\frac{\partial}{\partial t_1}\left(
{\cal B}^0-\frac{\partial{\cal D}^0}{\partial t_2}
\right)\quad {\rm i.e.} \quad
\pi^0=0.
\end{equation}
(One can derive $\partial{\cal B}^0/\partial t_1$, subtract it from
${\cal A}^0$, and compare the result with $\partial/\partial
t_2({\cal C}^0-\partial{\cal D}^0/\partial t_1)$.)

Now, we calculate the tail
\begin{equation}\label{pi1al}
\pi_a^\alpha={\cal A}_a^\alpha-\frac{\partial{\cal B}_a^\alpha}{\partial
t_1}-
\frac{\partial{\cal C}_a^\alpha}{\partial t_2}+
\frac{\partial^2 {\cal D}_a^\alpha}{\partial t_1\partial t_2},
\end{equation}
where
\begin{eqnarray}\label{ABCK}
{\cal D}_a^\alpha&=&\frac{1}{2\pi}\int_0^{2\pi}{\rm d}\varphi
\frac{K_a^\alpha}{{\rm q}r_1r_2},\qquad
{\cal B}_a^\alpha=\frac{1}{2\pi}\int_0^{2\pi}{\rm d}\varphi
\frac{K_a^\alpha c_2}{{\rm q}r_1(r_2)^2},\\
{\cal C}_a^\alpha&=&\frac{1}{2\pi}\int_0^{2\pi}{\rm
d}\varphi\frac{K_a^\alpha c_1}{{\rm q}(r_1)^2r_2},
\qquad
{\cal A}_a^\alpha=\frac{1}{2\pi}\int_0^{2\pi}{\rm d}\varphi
\frac{K_a^\alpha c_1c_2}{{\rm q}(r_1)^2(r_2)^2}.
\nonumber
\end{eqnarray}
The zeroth component, $K_a^0=k_a^0$, does not depend on $\varphi$.
Inserting relations ${\cal D}_a^0=k_a^0{\cal D}^0$, ${\cal
B}_a^0=k_a^0{\cal B}^0$, ${\cal C}_a^0=k_a^0{\cal C}^0$, and ${\cal
A}_a^0=k_a^0{\cal A}^0$ into eq.(\ref{pi1al}) and taking into account
identity (\ref{pi0}) yields
\begin{equation}\label{piK0}
\pi_1^0={\cal B}^0-\frac{\partial {\cal D}^0}{\partial t_2},\quad
\pi_2^0={\cal C}^0-\frac{\partial {\cal D}^0}{\partial t_1}.
\end{equation}

Space components, $K_a^i$, depend on $\varphi$. They can be expressed as
the scalar product $({\sf K}_a^i\cdot n)$ where components of three-vectors
${\sf K}_a^i\in V$ are as follows:
\begin{equation}
{\sf K}_{a,0}^i=n_{\rm q}^ik_a^3,\quad {\sf
K}_{a,1}^i=h\omega^i{}_1,\quad {\sf K}_{a,2}^i=h\omega^i{}_2.
\end{equation}
Here $\omega^i{}_j$ are components of the orthogonal matrix (\ref{om}).
Having substituted ${\sf K}_a^i$ for ${\sf a}$ in expressions (\ref{Dfin}),
(\ref{Bfin}), and (\ref{BA}) we obtain the terms ${\cal D}_a^i$, ${\cal
B}_a^i$ and ${\cal A}_a^i$, respectively. The last term, ${\cal C}_a^i$,
can be obtained from ${\cal B}_a^i$ via reciprocity.
To differentiate them we need the equalities
\begin{eqnarray}
\frac{\partial ({\sf K}_1^i\cdot {\sf r}_1)}{\partial t_1}&=&
({\sf K}_1^i\cdot {\sf c}_1) -v_1^i{\sf r}_1^0 -
\frac{({\mathbf n}_{\rm q}{\mathbf v}_1)}{{\rm q}}({\sf K}_1^i\cdot {\sf
r}_1)
-\frac{n_{\rm q}^i}{{\rm q}}({\sf r}_1\cdot {\sf r}_1)\nonumber\\
&-&\frac{k_2^3}{{\rm q}h^2}\left[
{\sf K}_{1,0}^i({\sf r}_1\cdot {\sf r}_1)+
{\sf r}_1^0({\sf K}_1^i\cdot {\sf r}_1)\right],\nonumber\\
\frac{\partial ({\sf K}_1^i\cdot {\sf r}_2)}{\partial t_1}&=&
-v_1^i{\sf r}_2^0-
\frac{({\mathbf n}_{\rm q}{\mathbf v}_2)}{{\rm q}}({\sf K}_1^i\cdot {\sf
r}_1)
-\frac{n_{\rm q}^i}{{\rm q}}({\sf r}_1\cdot {\sf r}_2)\nonumber\\
&+&\frac{k_2^3}{{\rm q}h^2}\left[{\sf r}_2^0({\sf K}_1^i\cdot {\sf r}_1)-
{\sf K}_{1,0}^i({\sf r}_1\cdot {\sf r}_2)-
2{\sf r}_1^0({\sf K}_1^i\cdot {\sf r}_2)\right],\nonumber\\
\frac{\partial ({\sf K}_1^i\cdot {\sf r}_1)}{\partial t_2}&=&
\frac{({\mathbf n}_{\rm q}{\mathbf v}_1)}{{\rm q}}({\sf K}_1^i\cdot {\sf
r}_2)
+\frac{n_{\rm q}^i}{{\rm q}}({\sf r}_1\cdot {\sf r}_2)\nonumber\\
&+&\frac{k_1^3}{{\rm q}h^2}\left[-{\sf r}_1^0({\sf K}_1^i\cdot {\sf r}_2)+
{\sf K}_{1,0}^i({\sf r}_1\cdot {\sf r}_2)+
2{\sf r}_2^0({\sf K}_1^i\cdot {\sf r}_1)\right],\nonumber\\
\frac{\partial ({\sf K}_1^i\cdot {\sf r}_2)}{\partial t_2}&=&
({\sf K}_1^i\cdot {\sf c}_2) +
\frac{({\mathbf n}_{\rm q}{\mathbf v}_2)}{{\rm q}}({\sf K}_1^i\cdot {\sf
r}_2)
+\frac{n_{\rm q}^i}{{\rm q}}({\sf r}_2\cdot {\sf r}_2)\nonumber\\
&+&\frac{k_1^3}{{\rm q}h^2}\left[
{\sf K}_{1,0}^i({\sf r}_2\cdot {\sf r}_2)+
{\sf r}_2^0({\sf K}_1^i\cdot {\sf r}_2)\right],\nonumber
\end{eqnarray}
in addition to eqs.(\ref{dra}) and (\ref{d1ra}).

The derivation of equalities
\begin{eqnarray}\label{DBC1i}
{\cal C}_1^i-\frac{\partial {\cal D}_1^i}{\partial t_1}&=&v_1^i{\cal D}^0
-\frac{n_{\rm q}^i}{{\rm q}^2\|{\sf r}_2\|}-
\frac{1}{{\rm q}\|{\sf r}_2\|^3}\frac{({\mathbf n}_{\rm q}{\mathbf
v}_2)}{{\rm q}}
({\sf K}_1^i\cdot {\sf r}_2)\\
&+&\frac{1}{{\rm q}\|{\sf r}_2\|^3}\frac{k_2^3}{{\rm q}}
\left(
{\sf K}_{1,0}^i[{\mathbf n}_{\rm q}{\mathbf v}_2]^2+
{\sf r}_2^0[{\mathbf n}_{\rm q}[{\mathbf v}_2{\mathbf n}_{\rm q}]]^i
\phantom{\frac11}\!\!\!\!\right),\nonumber
\\
{\cal B}_1^i-\frac{\partial {\cal D}_1^i}{\partial t_2}&=&
\frac{n_{\rm q}^i}{{\rm q}^2\|{\sf r}_1\|}+
\frac{1}{{\rm q}\|{\sf r}_1\|^3}\frac{({\mathbf n}_{\rm q}{\mathbf
v}_1)}{{\rm q}}
({\sf K}_1^i\cdot {\sf r}_1)\nonumber\\
&-&\frac{1}{{\rm q}\|{\sf r}_1\|^3}\frac{k_1^3}{{\rm q}}
\left(
{\sf K}_{1,0}^i[{\mathbf n}_{\rm q}{\mathbf v}_1]^2+
{\sf r}_1^0[{\mathbf n}_{\rm q}[{\mathbf v}_1{\mathbf n}_{\rm q}]]^i
\phantom{\frac11}\!\!\!\!\right),\nonumber
\end{eqnarray}
is virtually identical to that presented above, and we shall not bother
with the details. By virtue of the relation $K_2^i-K_1^i=q^i$ the
integrals (\ref{ABCK}) are related as follows:
\begin{equation}
{\cal D}_2^i={\cal D}_1^i+q^i{\cal D}^0,\quad
{\cal B}_2^i={\cal B}_1^i+q^i{\cal B}^0,\quad
{\cal C}_2^i={\cal C}_1^i+q^i{\cal C}^0,\quad
{\cal A}_2^i={\cal A}_1^i+q^i{\cal A}^0.
\end{equation}
Equipped with these expressions we find
\begin{eqnarray}\label{DBC2i}
{\cal C}_2^i-\frac{\partial {\cal D}_2^i}{\partial t_1}&=&
-\frac{n_{\rm q}^i}{{\rm q}^2\|{\sf r}_2\|}-
\frac{1}{{\rm q}\|{\sf r}_2\|^3}\frac{({\mathbf n}_{\rm q}{\mathbf
v}_2)}{{\rm q}}
({\sf K}_2^i\cdot {\sf r}_2)\\
&+&\frac{1}{{\rm q}\|{\sf r}_2\|^3}\frac{k_2^3}{{\rm q}}
\left(
{\sf K}_{2,0}^i[{\mathbf n}_{\rm q}{\mathbf v}_2]^2+
{\sf r}_2^0[{\mathbf n}_{\rm q}[{\mathbf v}_2{\mathbf n}_{\rm q}]]^i
\phantom{\frac11}\!\!\!\!\right),\nonumber
\\
{\cal B}_2^i-\frac{\partial {\cal D}_2^i}{\partial t_2}&=&
v_2^i{\cal D}^0+\frac{n_{\rm q}^i}{{\rm q}^2\|{\sf r}_1\|}+
\frac{1}{{\rm q}\|{\sf r}_1\|^3}\frac{({\mathbf n}_{\rm q}{\mathbf
v}_1)}{{\rm q}}
({\sf K}_2^i\cdot {\sf r}_1)\nonumber\\
&-&\frac{1}{{\rm q}\|{\sf r}_1\|^3}\frac{k_1^3}{{\rm q}}
\left(
{\sf K}_{2,0}^i[{\mathbf n}_{\rm q}{\mathbf v}_1]^2+
{\sf r}_1^0[{\mathbf n}_{\rm q}[{\mathbf v}_1{\mathbf n}_{\rm q}]]^i
\phantom{\frac11}\!\!\!\!\right).\nonumber
\end{eqnarray}
Finally, after a straightforward (but fairly lengthy) calculations
we derive the following relations:
\begin{equation}\label{pii}
\pi_1^i=v_1^i\left({\cal B}^0-\frac{\partial{\cal D}^0}{\partial
t_2}\right),\qquad \pi_2^i=v_2^i\left({\cal C}^0-\frac{\partial{\cal
D}^0}{\partial t_1}\right),
\end{equation}
which generalize eqs.(\ref{piK0}).

We will need also the tail
\begin{equation}\label{pialb}
\pi_{12}^{\alpha\beta}=
\frac{\partial^2 {\cal D}_{12}^{\alpha\beta}}{\partial t_1\partial t_2}-
\frac{\partial{\cal B}_{12}^{\alpha\beta}}{\partial t_1}-
\frac{\partial{\cal C}_{12}^{\alpha\beta}}{\partial t_2}+{\cal
A}_{12}^{\alpha\beta},
\end{equation}
where
\begin{eqnarray}\label{ABCKK}
{\cal
D}_{12}^{\alpha\beta}&=&\frac{\displaystyle 1}{\displaystyle
2\pi}\int_0^{2\pi}{\rm d}\varphi \frac{K_1^\alpha K_2^\beta}{{\rm
q}r_1r_2},\qquad
{\cal B}_{12}^{\alpha\beta}=\frac{1}{2\pi}\int_0^{2\pi}{\rm d}\varphi
\frac{K_1^\alpha K_2^\beta c_2}{{\rm q}r_1(r_2)^2},\\
{\cal C}_{12}^{\alpha\beta}&=&\frac{1}{2\pi}\int_0^{2\pi}{\rm d}\varphi
\frac{K_1^\alpha K_2^\beta c_1}{{\rm q}(r_1)^2r_2},
\qquad
{\cal A}_{12}^{\alpha\beta}=\frac{1}{2\pi}\int_0^{2\pi}{\rm d}\varphi
\frac{K_1^\alpha K_2^\beta c_1c_2}{{\rm q}(r_1)^2(r_2)^2}.
\nonumber
\end{eqnarray}
It can be obtained by means of covariant generalization of previous
relations. Setting $\alpha=0$ and $\beta=0$ and taking into account
eq.(\ref{pi0}), we obtain
\begin{eqnarray}\label{pi00}
\pi_{12}^{00}&=&k_1^0\left({\cal C}^0-\frac{\partial {\cal D}^0}{\partial
t_1}\right)+
k_2^0\left({\cal B}^0-\frac{\partial {\cal D}^0}{\partial t_2}\right)
+{\cal D}^0\nonumber\\
&=&{\cal C}_1^0-\frac{\partial {\cal D}_1^0}{\partial t_1}+
{\cal B}_2^0-\frac{\partial {\cal D}_2^0}{\partial t_2}-{\cal D}^0,
\end{eqnarray}
where relations ${\cal D}_a^0=k_a^0{\cal D}^0$, ${\cal
B}_a^0=k_a^0{\cal B}^0$, and ${\cal C}_a^0=k_a^0{\cal C}^0$ are
taken into account. If $\alpha=i$ and $\beta=0$, we have
\begin{eqnarray}\label{pii0}
\pi_{12}^{i0}&=&{\cal C}_1^i-\frac{\partial {\cal D}_1^i}{\partial
t_1}+
k_2^0v_1^i\left({\cal B}^0-\frac{\partial {\cal D}^0}{\partial t_2}\right)
\nonumber\\
&=&{\cal C}_1^i-\frac{\partial {\cal D}_1^i}{\partial t_1}+v_1^i\left(
{\cal B}_2^0-\frac{\partial {\cal D}_2^0}{\partial t_2}\right)
-v_1^i{\cal D}^0.
\end{eqnarray}
If $\alpha=0$ and $\beta=j$, we arrive at
\begin{eqnarray}\label{pi0j}
\pi_{12}^{0j}&=&k_1^0v_2^j\left({\cal C}^0-\frac{\partial {\cal
D}^0}{\partial t_1}\right)+
{\cal B}_2^j-\frac{\partial {\cal D}_2^j}{\partial t_2}
\nonumber\\
&=&v_2^j\left({\cal C}_1^0-\frac{\partial {\cal D}_1^0}{\partial
t_1}\right)+
{\cal B}_2^j-\frac{\partial {\cal D}_2^j}{\partial t_2}
-v_2^j{\cal D}^0.
\end{eqnarray}
An obvious generalization of expressions (\ref{pi00})-(\ref{pi0j}) is
\begin{equation}\label{pi_alb}
\pi_{12}^{\alpha\beta}=
v_1^\alpha\left({\cal B}_2^\beta-\frac{\partial {\cal D}_2^\beta}{\partial
t_2}\right)
+v_2^\beta\left({\cal C}_1^\alpha-\frac{\partial {\cal
D}_1^\alpha}{\partial t_1}\right)-v_1^\alpha v_2^\beta {\cal D}^0.
\end{equation}

Since the angular integration leads to a combination of partial derivatives
in time variables, the end points are valuable only. At these points the
radius of circle $C(O,h)=S_1\cap S_2$ vanishes. Hence we can restrict
oneselves to calculation of expressions
${\cal B}_{12}^{ij}-\partial_2{\cal D}_{12}^{ij}$ and
${\cal C}_{12}^{ij}-\partial_1{\cal D}_{12}^{ij}$ at points where $h=0$.

To simplify the calculations as much as possible we express the
integrands of eqs.(\ref{ABCKK}) in form of expansions in powers of
$h$. It allows us to remove harmonic functions from denominators.
Since the derivatives $\partial h^2/\partial t_a$ do not vanish
whenever $h^2=0$ (see eq.(\ref{h2})), it is sufficient to expand
${\cal D}_{12}^{ij}$ up to the first order of this parameter,
\begin{eqnarray}
{\cal D}_{12}^{ij}&=&\frac{k_1^3k_2^3n_{\rm q}^in_{\rm q}^j}{{\rm
q}r_1^0r_2^0}\left[
1+\frac{h^2}{2}\left(
\frac{[{\mathbf n}_{\rm q}{\mathbf v}_1]^2}{(r_1^0)^2}+
\frac{[{\mathbf n}_{\rm q}{\mathbf v}_2]^2}{(r_2^0)^2}+
\frac{([{\mathbf n}_{\rm q}{\mathbf v}_1][{\mathbf n}_{\rm q}{\mathbf
v}_2])}{r_1^0r_2^0}
\right)
\right]\\
&+&\frac{h^2}{2}\frac{k_1^3n_{\rm q}^i[{\mathbf n}_{\rm q}[{\mathbf
v}_1{\mathbf n}_{\rm q}]]^j+k_2^3n_{\rm q}^j[{\mathbf n}_{\rm q}[{\mathbf
v}_1{\mathbf n}_{\rm q}]]^i}{{\rm q}(r_1^0)^2r_2^0}\nonumber\\
&+&\frac{h^2}{2}\frac{k_1^3n_{\rm q}^i[{\mathbf n}_{\rm q}[{\mathbf
v}_2{\mathbf n}_{\rm q}]]^j+k_2^3n_{\rm q}^j[{\mathbf n}_{\rm q}[{\mathbf
v}_2{\mathbf n}_{\rm q}]]^i}{{\rm q}r_1^0(r_2^0)^2}\nonumber\\
&+&\frac{h^2}{2}\frac{\delta^{ij}-n_{\rm q}^in_{\rm q}^j}{{\rm
q}r_1^0r_2^0}.\nonumber
\end{eqnarray}
With degree of accuracy sufficient for our purposes,
\begin{equation}
{\cal B}_{12}^{ij}=\frac{n_{\rm q}^ik_1^3n_{\rm q}^jk_2^3{\sf
c}_{2,0}}{{\rm q}r_1^0(r_2^0)^2},
\qquad
{\cal C}_{12}^{ij}=\frac{n_{\rm q}^ik_1^3n_{\rm q}^jk_2^3{\sf
c}_{1,0}}{{\rm q}(r_1^0)^2r_2^0}.
\end{equation}
The calculation is straightforward, although it involves a fair
amount of algebra. Finally, we obtain
\begin{eqnarray}\label{DBC12}
{\cal B}_{12}^{ij}-\frac{\partial {\cal D}_{12}^{ij}}{\partial t_2}&=&
k_1^3n_{\rm q}^i\left(
{\cal B}_2^j-\frac{\partial {\cal D}_2^j}{\partial t_2}
\right)
+k_2^3n_{\rm q}^j\left(
{\cal B}_1^i-\frac{\partial {\cal D}_1^i}{\partial t_2}
\right)\\
&-&n_{\rm q}^ik_1^3n_{\rm q}^jk_2^3\left(
{\cal B}^0-\frac{\partial {\cal D}^0}{\partial t_2}
\right)-k_1^3\frac{\delta^{ij}-n_{\rm q}^in_{\rm q}^j}{{\rm
q}^2r_1^0};\nonumber\\
{\cal C}_{12}^{ij}-\frac{\partial {\cal D}_{12}^{ij}}{\partial t_1}&=&
k_2^3n_{\rm q}^j\left(
{\cal C}_1^i-\frac{\partial {\cal D}_1^i}{\partial t_1}
\right)
+k_1^3n_{\rm q}^i\left(
{\cal C}_2^j-\frac{\partial {\cal D}_2^j}{\partial t_1}
\right)\nonumber\\
&-&n_{\rm q}^ik_1^3n_{\rm q}^jk_2^3\left(
{\cal C}^0-\frac{\partial {\cal D}^0}{\partial t_1}
\right)+k_2^3\frac{\delta^{ij}-n_{\rm q}^in_{\rm q}^j}{{\rm
q}^2r_2^0}.\nonumber
\end{eqnarray}

\renewcommand{\thesubsubsection}{Appendix \Alph{subsubsection}}
\subsubsection{Bound parts of mixed energy-momentum and
angular momentum}\label{bound_i}
\renewcommand{\theequation}{\Alph{subsubsection}.\arabic{equation}}
\setcounter{equation}{0}

In this Appendix we consider the parts of mixed energy and mixed
momentum which describe unavoidable deformations of electromagnetic
``clouds'' of charged particles due to mutual interaction. The
short-range terms will be absorbed by four-momenta of ``bare''
particles within the renormalization procedure. They arise from the
bound part of mixed momentum,
\begin{eqnarray} \label{Pibnd}
{\cal P}^i_{\rm int, bnd}&=&{\cal P}^i_{\rm int}-{\cal P}^i_{\rm int, rad}\\
&=&\frac{e_1e_2}{2}\left[
{\hat\Pi }_2^i\left(\frac{\partial\sigma}{\partial t_2}\right)
+{\hat\Pi }^0\left(v_2^i\lambda_1\right)
-\frac{\partial}{\partial t_1}\left(
v_2^i\frac{\partial\sigma}{\partial t_2}{\cal D}^0
\right)\right.\nonumber\\
&+&\left.{\hat\Pi }_1^i\left(\frac{\partial\sigma}{\partial t_1}\right)
+{\hat\Pi }^0\left(v_1^i\lambda_2\right)
-\frac{\partial}{\partial t_2}\left(
v_1^i\frac{\partial\sigma}{\partial t_1}{\cal D}^0
\right)\right],\nonumber
\end{eqnarray}
which is equal to the difference of the total interference momentum
(\ref{t0if}) and the radiative component (\ref{Pirad}).

Our next task is to integrate ${\cal P}^i_{\rm int, bnd}$ over time
variables by means of expressions (\ref{int1}) and (\ref{int2}). Recall
from Section \ref{mix} that it is sufficient to evaluate the arguments of
time differential operators at the ends of integration intervals.

\begin{figure}[h]
\begin{center}
\epsfclipon
\epsfig{file=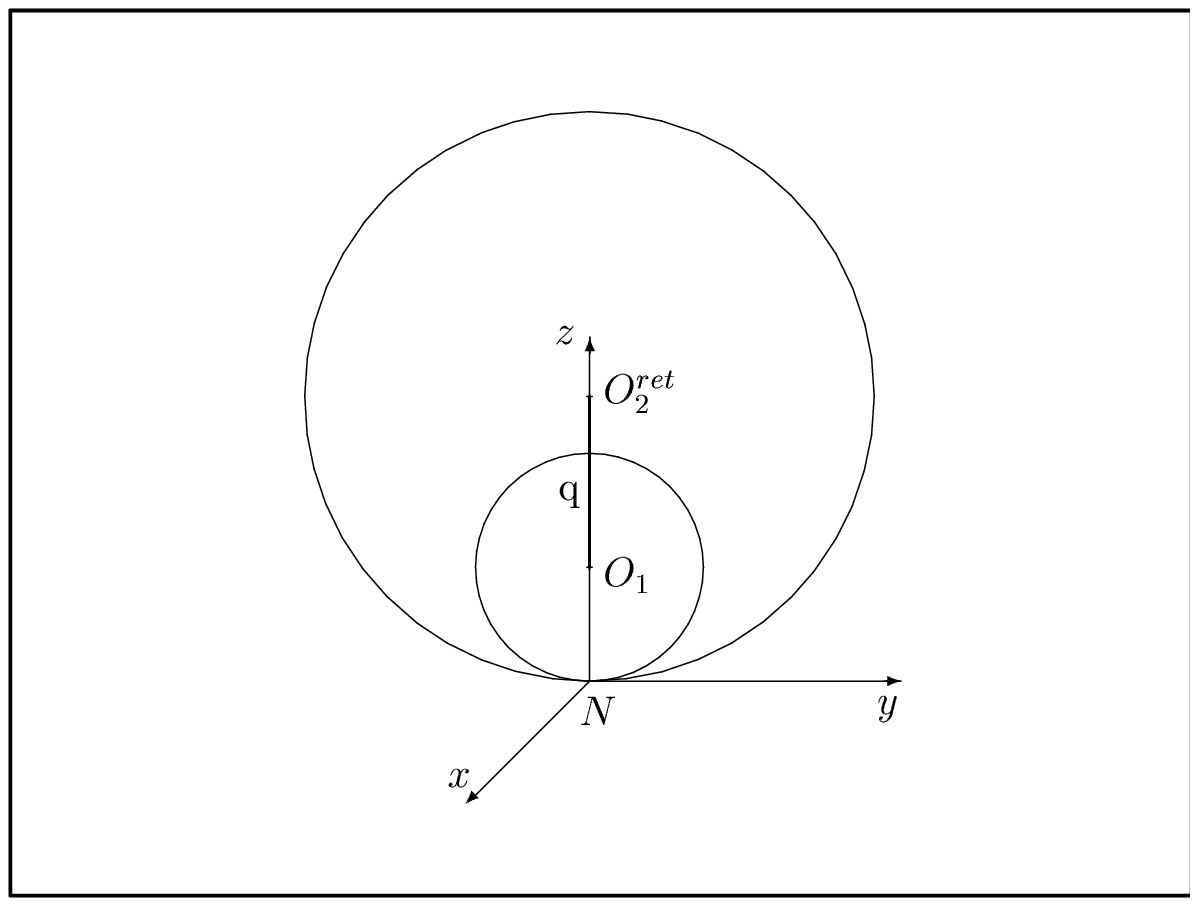,width=0.35\textwidth}$\qquad\qquad\qquad$
\epsfclipon
\epsfig{file=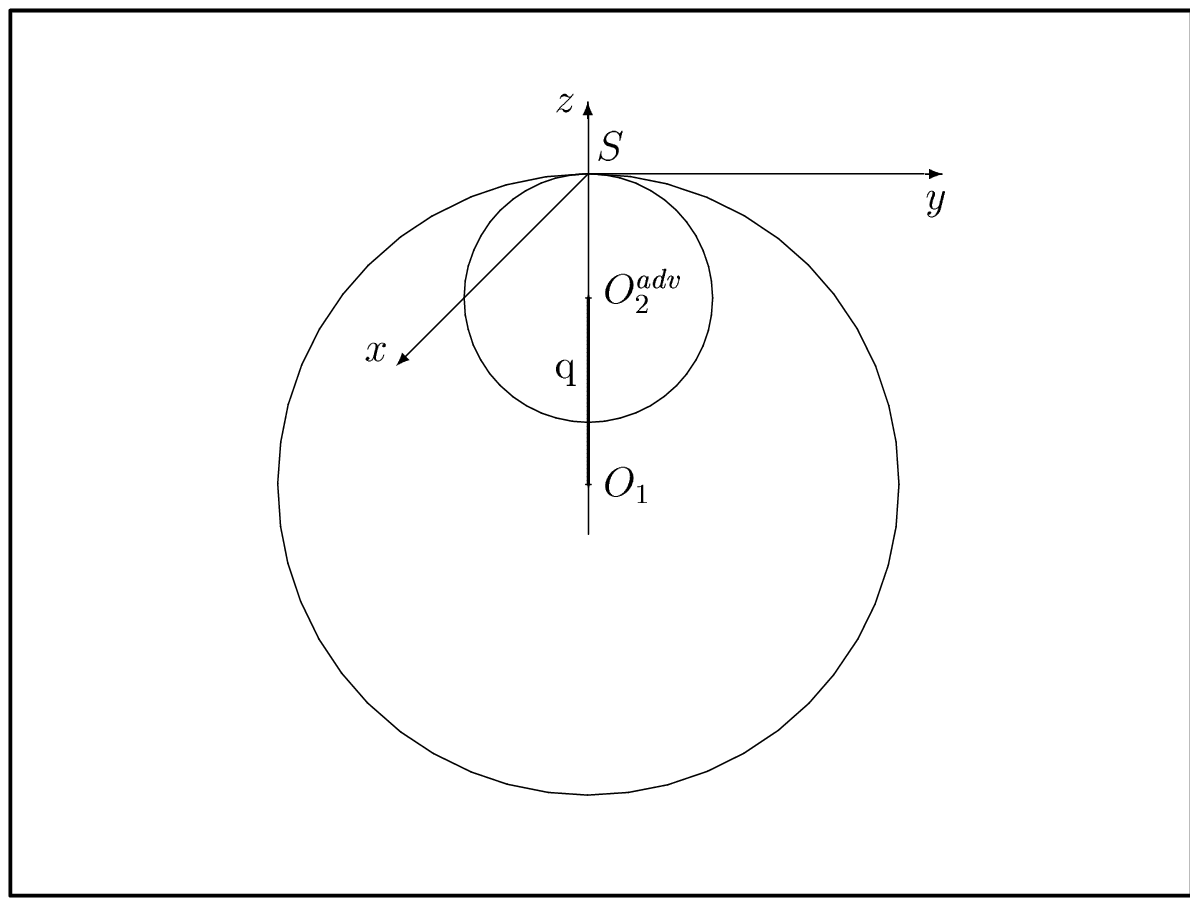,width=0.38\textwidth}
\end{center}
\caption{Boundary points of interference of spherical wave fronts
$S_1$ and $S_2$ in a hyperplane $\Sigma_t$. In momentarily comoving
Lorentz frame pictured in Fig.~\ref{kk} the wave fronts intersect at
the coordinate origin. If $h=0$, triangle $O_1O_2H$ reduces to the
line. Distance $\mathrm q$ between particles is equal to the
difference of radii of spheres.} \label{circles}
\end{figure}

In Figs.~\ref{circles} the boundary conditions of interference of
spherical wave fronts $S_1(O_1,k_1^0)$ and $S_2(O_2,k_2^0)$ are
presented. (Fig.~\ref{ra} pictures the combination of waves in
four-dimensional spacetime.) Distance ${\rm q}$ between their
centers, $O_1$ and $O_2$, is equal to the difference of their radii,
$k_1^0$ and $k_2^0$. In the left figure ${\rm q}=k_2^0-k_1^0$ while
in the right one ${\rm q}=k_1^0-k_2^0$. Inserting this into
eq.(\ref{k_k}) gives $k_a^3=k_a^0$ and $k_a^3=-k_a^0$ for the first
and the second cases, respectively.

The retarded and the advanced instants label the points on particles'
world lines which are connected by a null ray. Since the relative position
vector $q$ is of null length, the null vectors $K_1$ and $K_2$ are
collinear: $(K_1\cdot K_2)=0$. Their space parts ${\mathbf K}_1$ and
${\mathbf K}_2$ are codirectional in this case.

Within acausal region wave fronts combine in quite different manner
(see Fig.~\ref{ncirc}). In contrast to the pair of points $z_1(t_1)$
and $z_2[t_2^{ret}(t_1)]$, the vertices $z_1(t_1)\in\zeta_1$ and
$z_2(t_2')\in\zeta_2$ are spacelike related. Indeed, the scalar
product of the separation null vectors takes minimal value
$-2k_1^0k_2^0$ if their space parts ${\mathbf K}_1$ and ${\mathbf
K}_2$ are opposite directed. Hence $(q\cdot q)=4k_1^0k_2^0$.

\begin{figure}[t]
\begin{center}
\epsfclipon
\epsfig{file=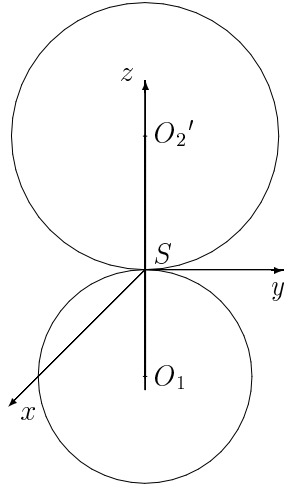,width=0.25\textwidth}
\end{center}
\caption{Acausal interference in a hyperplane $\Sigma_t$. Distance
$\mathrm q$ between particles placed at centers of spheres is equal to the
sum of their radii.} \label{ncirc}
\end{figure}

In Table \ref{krsl} we collect the basic quantities and functions which
are evaluated at the boundary points.

\begin{table}
\caption{Relative position, retarded distances and some basic functions at
the ends of integration intervals}\label{krsl}

\medskip

\begin{center}
\begin{tabular}{|c|c|c|}
\hline\hline
$[t_1,t_2^{ret}(t_1)]$, $[t_1^{adv}(t_2),t_2]$&
$[t_1^{ret}(t_2),t_2]$, $[t_1,t_2^{adv}(t_1)]$&
$[t_1,t_2'(t,t_1)]$, $[t_1'(t,t_2),t_2]$\\
\hline\hline
${\rm q}=k_2^0-k_1^0$, $q^0=+{\rm q}$&
${\rm q}=k_1^0-k_2^0$, $q^0=-{\rm q}$&
${\rm q}=k_1^0+k_2^0$\\
$k_1^3=k_1^0$, $k_2^3=k_2^0$&
$k_1^3=-k_1^0$, $k_2^3=-k_2^0$&
$k_1^3=-k_1^0$, $k_2^3=k_2^0$\\
$r_1=k_1^0\left[1-({\mathbf n}_q{\mathbf v}_1)\right]$&
$r_1=k_1^0\left[1+({\mathbf n}_q{\mathbf v}_1)\right]$&
$r_1=k_1^0\left[1+({\mathbf n}_q{\mathbf v}_1)\right]$\\
$r_2=k_2^0\left[1-({\mathbf n}_q{\mathbf v}_2)\right]$&
$r_2=k_2^0\left[1+({\mathbf n}_q{\mathbf v}_2)\right]$&
$r_2=k_2^0\left[1-({\mathbf n}_q{\mathbf v}_2)\right]$\\
\hline
$\sigma=0$&$\sigma=0$&$\sigma=-2k_1^0k_2^0$\\
$\frac{\displaystyle\partial\sigma}{\displaystyle\partial
t_1}={\rm q}\left[1-({\mathbf n}_q{\mathbf v}_1)\right]$&
$\frac{\displaystyle\partial\sigma}{\displaystyle\partial
t_1}=-{\rm q}\left[1+({\mathbf n}_q{\mathbf v}_1)\right]$&
$\frac{\displaystyle\partial\sigma}{\displaystyle\partial
t_1}=-{\rm q}\left[1+({\mathbf n}_q{\mathbf v}_1)\right]+2k_2^0$\\[1.5ex]
$\frac{\displaystyle\partial\sigma}{\displaystyle\partial
t_2}=-{\rm q}\left[1-({\mathbf n}_q{\mathbf v}_2)\right]$&
$\frac{\displaystyle\partial\sigma}{\displaystyle\partial
t_2}={\rm q}\left[1+({\mathbf n}_q{\mathbf v}_2)\right]$&
$\frac{\displaystyle\partial\sigma}{\displaystyle\partial
t_2}=-{\rm q}\left[1-({\mathbf n}_q{\mathbf v}_2)\right]+2k_1^0$\\
\hline
$\lambda_1=k_1^0{\rm q}\left[1-({\mathbf n}_q{\mathbf v}_1)\right]$&
$\lambda_1=-k_1^0{\rm q}\left[1+({\mathbf n}_q{\mathbf
v}_1)\right]$&
$\lambda_1=-k_1^0{\rm q}\left[1+({\mathbf n}_q{\mathbf
v}_1)\right]$\\
$\lambda_2=-k_2^0{\rm q}\left[1-({\mathbf n}_q{\mathbf v}_2)\right]$&
$\lambda_2=k_2^0{\rm q}\left[1+({\mathbf n}_q{\mathbf
v}_2)\right]$&
$\lambda_2=-k_2^0{\rm q}\left[1-({\mathbf n}_q{\mathbf
v}_2)\right]$\\
\hline
\end{tabular}
\end{center}
\end{table}

To integrate the expression (\ref{Pibnd}) over time variables, we
apply the scheme developed in Section \ref{mix}. Usage of the rule
(\ref{d1G0}) implies
\begin{eqnarray}
p_{{\rm bnd},21}^i&=&\left.
\frac{e_1e_2}{2}\frac{v_2^i+n_q^i}{k_2^0\left[1-({\mathbf
n}_q{\mathbf v}_2)\right]}\right|_{t_1\to-\infty}^{[t,t_2^{ret}(t)]}
+
\left.\frac{e_1e_2}{2}\frac{v_2^i-n_q^i}{k_2^0\left[1+({\mathbf
n}_q{\mathbf v}_2)\right]}\right|_{t_1\to-\infty}^{[t_1^{ret}(t),t]}\\
&+&
\left.\frac{e_1e_2}{2}\frac{v_2^i-n_q^i}{k_2^0\left[1-({\mathbf
n}_q{\mathbf v}_2)\right]}\right|^{[t,t_2^{ret}(t)]}_{[t_1^{ret}(t),t]}.
\nonumber
\end{eqnarray}
If one prefer another order of differentiation (\ref{d2G0}) they obtain
\begin{eqnarray}
p_{{\rm bnd},12}^i&=&\left.
\frac{e_1e_2}{2}\frac{v_1^i+n_q^i}{k_1^0\left[1-({\mathbf
n}_q{\mathbf v}_1)\right]}\right|_{t_2\to-\infty}^{[t,t_2^{ret}(t)]}
+
\left.\frac{e_1e_2}{2}\frac{v_1^i-n_q^i}{k_1^0\left[1+({\mathbf
n}_q{\mathbf v}_1)\right]}\right|_{t_2\to-\infty}^{[t_1^{ret}(t),t]}\\
&+&
\left.\frac{e_1e_2}{2}\frac{v_1^i+n_q^i}{k_1^0\left[1+({\mathbf
n}_q{\mathbf v}_1)\right]}\right|_{[t,t_2^{ret}(t)]}^{[t_1^{ret}(t),t]}.
\nonumber
\end{eqnarray}
The lower limits $t_a\to -\infty$ vanish even if the motion is
finite. Final expressions depend on particles' positions and
velocities referred to the moments $t_1^{ret}(t)$ and $t_2^{ret}(t)$
as well as on the laboratory time $t$ itself,
\begin{eqnarray}\label{pibnd21}
p_{{\rm bnd},21}^i&=&\left.e_1e_2\frac{v_2^i}{k_2^0\left[1-({\mathbf
n}_q{\mathbf v}_2)\right]}\right|^{[t,t_2^{ret}(t)]}
+
\left.\frac{e_1e_2}{2}\lim_{t_2\to
t}\frac{v_2^i-n_q^i}{k_2^0\left[1+({\mathbf n}_q{\mathbf
v}_2)\right]}\right|^{[t_1^{ret}(t_2),t_2]}\nonumber\\
&-&
\left.\frac{e_1e_2}{2}\lim_{t_2\to
t}\frac{v_2^i-n_q^i}{k_2^0\left[1-({\mathbf
n}_q{\mathbf v}_2)\right]}\right|_{[t_1'(t,t_2),t_2]};\\
p_{{\rm bnd},12}^i&=&\left.e_1e_2\frac{v_1^i}{k_1^0\left[1+({\mathbf
n}_q{\mathbf v}_1)\right]}\right|^{[t_1^{ret}(t),t]}
+
\left.\frac{e_1e_2}{2}\lim_{t_1\to
t}\frac{v_1^i+n_q^i}{k_1^0\left[1-({\mathbf n}_q{\mathbf
v}_1)\right]}\right|^{[t_1,t_2^{ret}(t_1)]}\nonumber\\
&-&
\left.\frac{e_1e_2}{2}\lim_{t_1\to
t}\frac{v_1^i+n_q^i}{k_1^0\left[1+({\mathbf
n}_q{\mathbf v}_1)\right]}\right|_{[t_1,t_2'(t,t_2)]}.\label{pibnd12}
\end{eqnarray}

In an analogous way we find short-range contribution to the mixed energy
due to time integration of the following expression:
\begin{eqnarray} \label{P0bnd}
{\cal P}^0_{\rm int, bnd}&=&{\cal P}^0_{\rm int}-{\cal P}^0_{\rm int, rad}\\
&=&e_1e_2\left[
{\hat\Pi }^0\left(k_1^0\frac{\partial\sigma}{\partial t_1}+
k_2^0\frac{\partial\sigma}{\partial t_2}+\sigma
-\frac12\frac{\partial\sigma}{\partial t_1}\frac{\partial\sigma}{\partial
t_2}\right)
\right.\nonumber\\
&-&\left.\frac12\frac{\partial}{\partial t_1}
\left(\frac{1}{{\rm q}\|{\sf r}_1\|}\frac{\partial^2\sigma}{\partial
t_1\partial t_2}\right)
-\frac12\frac{\partial}{\partial t_2}
\left(\frac{1}{{\rm q}\|{\sf r}_2\|}\frac{\partial^2\sigma}{\partial
t_1\partial t_2}\right)\right].\nonumber
\end{eqnarray}
The calculation is virtually identical to that presented above, and
we shall not bother with details. Finally, we obtain
\begin{eqnarray}\label{p0bnd21}
p_{{\rm bnd},21}^0&=&\left.\frac{e_1e_2}{k_2^0\left[1-({\mathbf
n}_q{\mathbf v}_2)\right]}\right|^{[t,t_2^{ret}(t)]}
+
\lim_{t_2\to t}\frac{e_1e_2}{k_2^0}\left[\frac{1}{1+({\mathbf
n}_q{\mathbf v}_2)}-1\right]^{[t_1^{ret}(t_2),t_2]},\\
p_{{\rm bnd},12}^0&=&\left.\frac{e_1e_2}{k_1^0\left[1+({\mathbf
n}_q{\mathbf v}_1)\right]}\right|^{[t_1^{ret}(t),t]}
+
\lim_{t_1\to t}\frac{e_1e_2}{k_1^0}\left[\frac{1}{1-({\mathbf
n}_q{\mathbf v}_1)}-1\right]^{[t_1,t_2^{ret}(t_1)]}.\label{p0bnd12}
\end{eqnarray}
As could be expected for bound terms, they
(i) depend on the momentary state of particles' motion,
(ii) contain divergent terms, and
(iii) are non-covariant.

The bound components of angular momentum have similar structure,
\begin{eqnarray}\label{Mbnd21}
M_{{\rm
bnd},21}^{0i}&=&e_1e_2\left[t\frac{v_2^i-n_q^i}{k_2^0\left[1-
({\mathbf n}_q{\mathbf v}_2)\right]}-\frac{z_2^i}{k_2^0}
+\frac{t_2n_q^i-z_2^in_q^0}{k_2^0\left[1-({\mathbf n}_q{\mathbf
v}_2)\right]}\right]^{[t,t_2^{ret}(t)]}\\
&+&
\frac{e_1e_2}{2}\lim_{t_2\to
t}\left[t\frac{v_2^i+n_q^i}{k_2^0\left[1+
({\mathbf n}_q{\mathbf v}_2)\right]}-\frac{z_2^i}{k_2^0}
-2\frac{t_2n_q^i-z_2^in_q^0}{k_2^0\left[1+({\mathbf n}_q{\mathbf
v}_2)\right]}\right]^{[t_1^{ret}(t),t]}\nonumber\\
&-&
\frac{e_1e_2}{2}\lim_{t_2\to
t}\left[t\frac{v_2^i-n_q^i}{k_2^0\left[1-
({\mathbf n}_q{\mathbf
v}_2)\right]}-\frac{z_2^i}{k_2^0}\right]^{[t_1'(t,t_2),t_2]},\nonumber\\
M_{{\rm
bnd},21}^{ij}&=&\left.e_1e_2\frac{z_1^iv_2^j-z_1^jv_2^i}{k_2^0\left[1-
({\mathbf n}_q{\mathbf v}_2)\right]}\right|^{[t,t_2^{ret}(t)]}\\
&-&\frac{e_1e_2}{2}\left.(n_q^iv_2^j-n_q^jv_2^i)\left(\frac{1}{1+({\mathbf
n}_q{\mathbf
v}_2)}+\frac{1}{1-({\mathbf
n}_q{\mathbf v}_2)}\right)\right|^{[t_1^{ret}(t),t]}\nonumber\\
&+&
\left.\frac{e_1e_2}{2}\lim_{t_2\to
t}\frac{z_2^i(v_2^j-n_q^j)-z_2^j(v_2^i-n_q^i)}{k_2^0\left[1+({\mathbf
n}_q{\mathbf v}_2)\right]}\right|^{[t_1^{ret}(t_2),t_2]}\nonumber\\
&-&
\left.\frac{e_1e_2}{2}\lim_{t_2\to
t}\frac{z_2^i(v_2^j-n_q^j)-z_2^j(v_2^i-n_q^i)}{k_2^0\left[1-({\mathbf
n}_q{\mathbf v}_2)\right]}\right|^{[t_1'(t,t_2),t_2]}.\nonumber
\end{eqnarray}
Alternative expressions, $M_{{\rm bnd},21}^{\mu\nu}$, can be obtained via
reciprocity of indices $1$ and $2$.

In contrast to one-particle case, expanding of the expressions under
limit signs in powers of $\Delta_a=t-t_a$ does not yield simple and
manifestly covariant terms of clear physical sense. The ``deformation''
is due to the choice of the coordinate-dependent hole around the particle
in the integration surface $\Sigma_t$. We neglect these structureless
terms.

\end{document}